\documentclass[reprint, amsmath, amssymb, superscriptaddress, pra]{revtex4-2}
\usepackage{graphicx, xcolor} 
\usepackage{hyperref}
\usepackage{bbm}
\usepackage{stmaryrd}

\newcommand*{\hham}{\hat{\mathcal{H}}}

\newcommand{\ket}[1]{\left|#1 \right\rangle }
\newcommand{\bra}[1]{\left\langle #1 \right|  }

\begin{document}
\title{Onset of transmon ionization in microwave single-photon detection}

\author{Yuki Nojiri}
\email{yuki.nojiri@wmi.badw.de}
\affiliation{Walther-Mei{\ss}ner-Institut, Bayerische Akademie der Wissenschaften, 85748 Garching, Germany}
\affiliation{Physics Department, School of Natural Sciences, Technical University of Munich, 85748 Garching, Germany}

\author{Kedar E. Honasoge}
\affiliation{Walther-Mei{\ss}ner-Institut, Bayerische Akademie der Wissenschaften, 85748 Garching, Germany}
\affiliation{Physics Department, School of Natural Sciences, Technical University of Munich, 85748 Garching, Germany}

\author{Achim Marx}
\affiliation{Walther-Mei{\ss}ner-Institut, Bayerische Akademie der Wissenschaften, 85748 Garching, Germany}

\author{Kirill G. Fedorov}
\email{kirill.fedorov@wmi.badw.de}
\author{Rudolf Gross}
\email{rudolf.gross@wmi.badw.de}
\affiliation{Walther-Mei{\ss}ner-Institut, Bayerische Akademie der Wissenschaften, 85748 Garching, Germany}
\affiliation{Physics Department, School of Natural Sciences, Technical University of Munich, 85748 Garching, Germany}
\affiliation{Munich Center for Quantum Science and Technology (MCQST), Schellingstr. 4, 80799 Munich, Germany}

\date{\today}
\begin{abstract}
By strongly driving a transmon-resonator system, the transmon qubit may eventually escape from its cosine-shaped potential. This process is called transmon ionization (TI) and  known to be detrimental to the qubit coherence and operation. In this work, we investigate the onset of TI in an irreversible, parametrically-driven, frequency conversion process in a system consisting of a superconducting 3D-cavity coupled to a fixed-frequency transmon qubit. Above a critical pump power we find a sudden increase in the transmon population. Using R{\'e}nyi entropy, Floquet modes, and Husimi Q functions, we infer that this abrupt change can be attributed to a quantum-to-classical phase transition. Furthermore, in the context of the single-photon detection, we measure a TI-uncorrected detection efficiency of up to 86\% and estimate a TI-corrected value of up to $78$\% by exploiting the irreversible frequency conversion. Our numerical simulations suggest that increasing the detuning between the pump and qubit frequencies and increasing the qubit anharmonicity can suppress the TI impact. Our findings highlight the general importance of the TI process when operating coupled qubit-cavity systems.
\end{abstract}

\maketitle

\section{Introduction}
In superconducting circuits and cavity quantum electrodynamics systems, Josephson junctions play a pivotal role. Characterized by their strong nonlinearity, and low losses, these junctions are fundamental for various quantum information applications. Their crucial role is evident in the realization of parametrically driven devices, including quantum amplifiers \cite{Yurke1989, Castellanos-Beltran2008, Bergeal2010, Abdo2014, Malnou2021, Qiu2023}, frequency converters \cite{Abdo2011, Kamal2014, Jiang2023}, nonclassical light generators \cite{Pechal2014, Flurin2012}, stabilizers \cite{Murch2012, Leghtas2015}, and microwave single-photon detectors (SPDs) \cite{Dassonneville2020,Lescanne2020}. This is attributed to three- or four-wave mixing processes enabled by a strong microwave drive, denoted as the "pump". Recently, SPDs have attracted particular attention as they can be applied for many quantum protocols, such as quantum teleportation \cite{Ma2012b}, entanglement distillation \cite{Abdelkhalek2016}, entanglement swapping \cite{Ma2012a}, quantum erasers \cite{Scully1982}, quantum repeaters \cite{Yuan2008}, Gaussian boson sampling \cite{Deng2023}, and quantum error correction \cite{Jin2012}.

The development of SPDs in the microwave domain is particularly challenging. Compared to devices operating in the optical regime, this is primarily attributed to the lower photon energy \cite{Chen2011, Inomata2016, Kono2018, Besse2018, Dassonneville2020, Lescanne2020} and the high performance standards demanded by applications in quantum information, quantum sensing, and quantum illumination \cite{Varnava2008, Hadfield2009, Lamoreaux2013, Chunnilall2014, Barzanjeh2015, Assouly2023, Kronowetter2023}. Despite these obstacles, various microwave SPD designs with quantum efficiencies exceeding 50\% have been demonstrated in recent years \cite{Chen2011, Inomata2016, Kono2018, Besse2018, Dassonneville2020, Lescanne2020}. The strategic significance of microwave SPDs becomes evident in applications such as dark matter axion search \cite{Lamoreaux2013}, spin fluorescence detection \cite{Wang2023, Billaud2023}, and quantum radar \cite{LasHeras2017,Assouly2023}. For example recently, the application of microwave SPDs in quantum radar systems demonstrated an experimental quantum advantage in a target detection \cite{Assouly2023}. In the field of spin fluorescence detection, using microwave SPDs allows one to surpass the signal-to-noise-ratio performance of conventional electron spin resonance methods \cite{Wang2023,Billaud2023}. Microwave SPDs are also expected to further advance the rapidly growing fields of microwave quantum communication and sensing \cite{Flurin2012, Fedorov2021, Renger2021, Kronowetter2023}.

\begin{figure*}[hbt]
	\centering
	\includegraphics[width=1.95 \columnwidth]{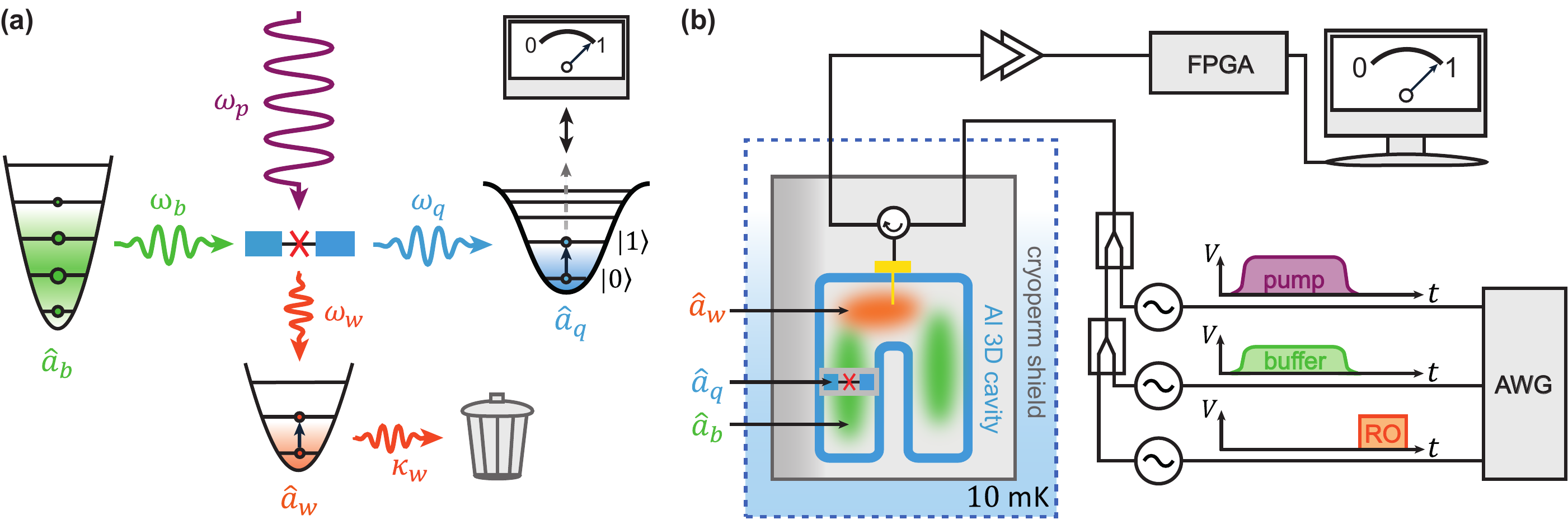}
	\caption{(a) Principle of the microwave single-photon detection. A coherent incoming photon (green) is absorbed by a buffer mode, $\hat{a}_b$, and is converted to a pair of qubit-waste excitations described with operators $\hat{a}_q$ and $\hat{a}_w$, respectively, with the interaction strength $g_4 \xi_p$. Due to the engineered fast dissipation of the waste mode, $\kappa_w\gg \left| g_4 \xi_p \right|$, the inverse process $(\hat{a}_b^\dagger\hat{a}_q\hat{a}_w)$ is effectively suppressed. (b) Schematic of the experimental setup. The transmon is mounted in one arm of the horseshoe-shaped 3D superconducting cavity and coupled to the cavity waste/buffer modes, represented by orange/green shaded areas depicting corresponding electric field distributions, respectively. The buffer (green), readout (orange), and strong pump pulses (purple) are shaped using an arbitrary waveform generator (AWG) for the time-domain experiments. The frequencies of the buffer, $\omega_b$, readout, $\omega_w$, and pump tone, $\omega_p$, are set by external microwave generators.}
	\label{fig:SPD_principle}
\end{figure*}

A common technique for detecting single photons by microwave SPDs involves the use of transmon qubits, the most common qubit type used in quantum computation \cite{Kono2018, Besse2018, Dassonneville2020, Lescanne2020}. The transmon ionization (TI) emerges when the qubit escapes from the Josephson potential by strong microwave driving, transferring the qubit into a complex regime involving many quantum levels \cite{Lescanne2019, Verney2019, Shillito2022, Cohen2023}. The phenomena occuring in this regime are known under various names, such as the quantum-to-classical phase transition \cite{Boissonneault2010, Bishop2010,Pietikainen2019}, first-order dissipative phase transition \cite{Carmichael2015, Mavrogordatos2016, Fink2018, Chen2023}, chaotic dynamics \cite{Utermann1994,Cohen2023}, and breakdown of photon-blockade \cite{Carmichael2015, Pietikainen2019}. Various tools and representations, such as the R{\'e}nyi entropy, Floquet theory, and Husimi Q functions, are employed to identify potential phase transitions. Notably, within the framework of the quantum Duffing oscillator, the R{\'e}nyi entropy serves as a quantifier of purity of the quantum system, $P=\text{Tr}\left( \hat{\rho}^2 \right)$, while the Floquet theory provides insights into the temporal evolution under periodic driving. Concurrently, the Husimi Q function offers a phase space representation, revealing significant quantum fluctuations which arise from tunneling between two metastable states. While this bistability is known to be useful for high-fidelity readout schemes \cite{Shillito2022,Reed2010}, its influence on the SPD behavior is not well understood so far. Given the importance of such devices in quantum applications, we investigate its complex dynamics, and quantify the impact of the TI on the SPD performance.

In the present work, we focus on the onset of TI in the irreversible frequency conversion process in strongly driven transmon-resonator systems applied for single-photon detection. The model and the detection principle of our device are explained in Sec.\,\ref{sec:model}. In Sec.\,\ref{sec:ionization}, we study the conversion of the incoming photons to the transmon excitations and waste resonator photons as a function of the pump power. Once the pump power eventually approaches a critical threshold, it triggers the sudden increase in the transmon population in response to the incoming buffer photons. We study this process by using the R{\'e}nyi entropy, the Floquet theory, and the Husimi Q function. Our results indicate that the abrupt change of the transmon response can be attributed to a quantum-to-classical phase transition. Subsequently in Sec.\,\ref{sec:spd_performance}, we provide a comprehensive analysis of the irreversible frequency conversion in our system, leading to a single-photon detection efficiency of $86$\%. Then, we consider the influence of TI on the detector performance and estimate the TI-corrected detection efficiency to be notably lower, around $78$\%. Concluding our study, we systematically evaluate strategies to enhance the SPD detection effi-ciency by properly adjusting the system parameters. In particular, we find that increasing the detuning between the pump and qubit frequencies and increasing the qubit anharmonicity are the most promising steps for further increasing the detection efficiency. Furthermore, our analysis suggests that the spectral arrangement of the buffer and waste frequencies should be implemented based on the sign of the qubit anharmonicity to ensure that the pump drive and multi-photon qubit transitions do not coincide in the frequency space.

\begin{figure*}[bt]
	\centering
	\includegraphics[width= 1.95\columnwidth]{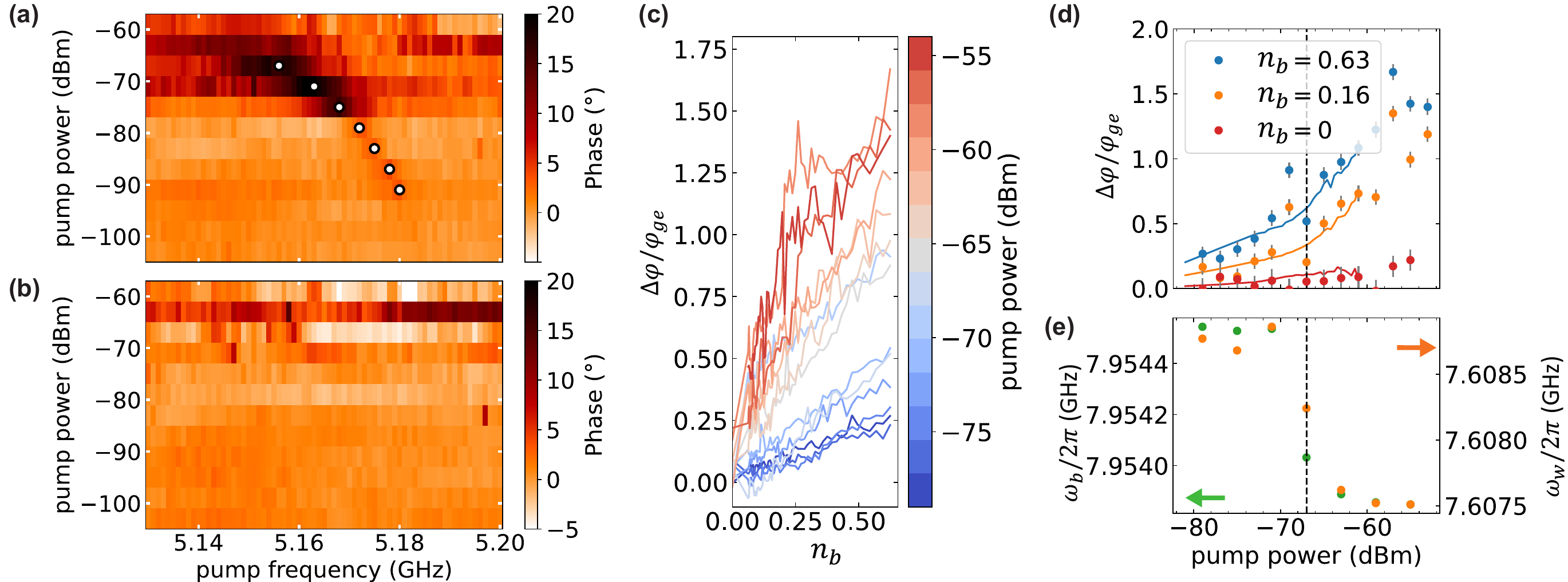}
	\caption{(a) Triple-tone pulsed spectroscopy of the coupled transmon-resonator system as explained in the main text. An inclined feature (white dots) indicates the four-wave mixing process. (b) Reference two-tone pulsed spectroscopy of the coupled transmon-resonator system without the buffer signal. (c) Phase shift of the transmon response $\Delta\varphi$ normalized to the phase shift between the ground and first excited states $\varphi_{ge}$ as a function of the steady state buffer photon number $n_b$. Different colors correspond to specific pump power values as indicated by the color code. (d)\,Transmon phase response for three selected buffer photon numbers $n_b = 0.00, 0.16, 0.63$ (red, orange, blue) as a function of the pump power. The vertical black dashed line indicates the CPP, at which the abrupt change of the qubit population can be observed. Solid lines are the simulation results using Eq.\,\eqref{eq:Ham4} with $n_b = 0.00, 0.16, 0.63$ (red, orange, blue). (e)\,The buffer and waste resonance frequencies as a function of the pump power at $\omega_p/2\pi = 5.1595$\,GHz. A distinct jump of the resonance frequencies can be observed in both resonators at the CPP of $-67$\,dBm.}
	\label{fig:nq_Pp}
\end{figure*}

\section{Demonstration of irreversible frequency conversion process}
\label{sec:model}
In our experiments, we use a horseshoe-shaped aluminum 3D cavity, which contains a transmon qubit chip (as schematically shown in Fig.\,\ref{fig:SPD_principle}(b)). We utilize two of the cavity modes: the waste mode at $\omega_w/2\pi = 7.609$\,GHz with a decay rate of $\kappa_w/2\pi = 16.7$\,MHz and the buffer mode at $\omega_b/2\pi = 7.955$\,GHz with a decay rate of $\kappa_b/2\pi = 3.7$\,MHz. The qubit frequency is $\omega_q/2\pi = 5.664$\,GHz, with $T_1 =28 \ \mathrm \mu$s and $T_2 = 16 \ \mathrm \mu$s, limited by the Purcell effect. The qubit is coupled to the waste and buffer modes with coupling strengths of $g_w/2\pi = 30$\,MHz and $g_b/2\pi = 18$\,MHz, respectively. In a driven system, the ac-Stark effect induces shifts in all three eigenfrequencies as a function of the pump power. These shifted frequencies are denoted by $\omega'_j$ for the $j$-th mode, where $j = \{q, b, w\}$. The Hamiltonian for the irreversible frequency conversion process in the rotating frame of $\hham_0/\hbar = \omega'_q \hat{a}_q^\dagger \hat{a}_q +  \omega'_w\hat{a}_w^\dagger\hat{a}_w + \omega'_b\hat{a}_b^\dagger\hat{a}_b$ is formulated as \cite{Lescanne2019}
\begin{align}
	\hham_4/\hbar& \approx\sum_{k=2}^{N_t-1}\frac{\chi^{(k)}}{k!} \left(\hat{a}_q^\dagger\right)^k \hat{a}_q^k\nonumber\\
    & + \chi_{qw} \hat{a}_q^\dagger \hat{a}_q \hat{a}_w^\dagger \hat{a}_w + \chi_{qb} \hat{a}_q^\dagger \hat{a}_q \hat{a}_b^\dagger  \hat{a}_b + i\varepsilon_b\left( \hat{a}_b^\dagger - \hat{a}_b \right) \nonumber\\
	& + g_{4} \left(\xi_p  \hat{a}_q^\dagger \hat{a}_w^\dagger\hat{a}_b e^{i\Delta_{qwbp}t}+ \xi_p ^* \hat{a}_q\hat{a}_w\hat{a}_b^\dagger e^{-i\Delta_{qwbp}t} \right)  \nonumber\\
	& + 2i\varepsilon_q \cos\left(\omega_pt\right)\left( \hat{a}_q^\dagger e^{i\omega'_q t} - \hat{a}_q e^{-i\omega'_q t} \right),
	\label{eq:Ham4}
\end{align}
where $\hat{a}_q$, $\hat{a}_w$, and $\hat{a}_b$ are the annihilation operators of the qubit, waste, and buffer modes, respectively, and $N_t$ is the dimension of the transmon Hilbert space. The detuning involving all four frequencies is defined as $\Delta_{qwbp}=\omega'_q+\omega'_w-\omega'_b-\omega_p$ and the driving strength of the transmon and buffer mode as $\varepsilon_{q,b}$, respectively. The term $\chi^{(k)}$ corrects for the $k$-th eigenfrequency of the transmon, with $\chi^{(2)}$ being its anharmonicity. In addition, $\chi_{qw}$ and $\chi_{qb}$ are the cross-Kerr interaction strengths of the waste and buffer modes, respectively. Lastly, $g_{4}$ is the strength of the four-wave mixing interaction, and $\xi_p$ is the pump amplitude at the pump frequency, $\omega_p$. Below the regime, where $\chi^{(2)}\gg \varepsilon_{q,b}, \kappa_{b,w}, 1/T_{1,2}$, the bosonic annihilation operator $\hat{a}_q$ can be replaced by the Pauli operator $\hat{\sigma}$. The irreversible frequency conversion process is based on the dissipation-engineered four-wave mixing process of a nonlinear Josephson junction element \cite{Lescanne2020}. When an incoming photon arrives at the buffer mode frequency, the buffer photon is converted to the excited state of the qubit and a waste photon by a strong pump pulse, corresponding to the parametrically activated conversion process $(\hat{a}_b\hat{\sigma}^\dagger\hat{a}_w^\dagger)$. However, under the conditions, $\kappa_w \gg \left| g_4 \xi_p \right| $, the waste resonator state rapidly decays to the vacuum state, making the process irreversible and prohibiting the inverse process $(\hat{a}_b^\dagger\hat{\sigma}\hat{a}_w)$. As a result, information of the incoming photon is stored in the qubit state, as shown in Fig.\,\ref{fig:SPD_principle}\,(a).

By performing a triple-tone spectroscopy composed of the pump and buffer pulses with identical duration of $t_b = 20$\,$\mu$s, followed by a readout pulse at the waste mode frequency with a duration of $t_r = 2.5$\,$\mu$s (see Fig.\,\ref{fig:SPD_principle}), we observe the conversion process of the buffer photons to the qubit excited state. As shown in Fig.\,\ref{fig:nq_Pp}\,(a), when the buffer pulse is switched on, we observe the phase response of the transmon in the pump frequency range of $\omega_p/2\pi=5.15-5.18$\,GHz. This response disappears in the absence of the buffer signal (see Fig.\,\ref{fig:nq_Pp}\,(b)).

\section{onset of transmon ionization}
\label{sec:ionization}

\begin{figure*}[bt]
	\centering
	\includegraphics[width=1.9\columnwidth]{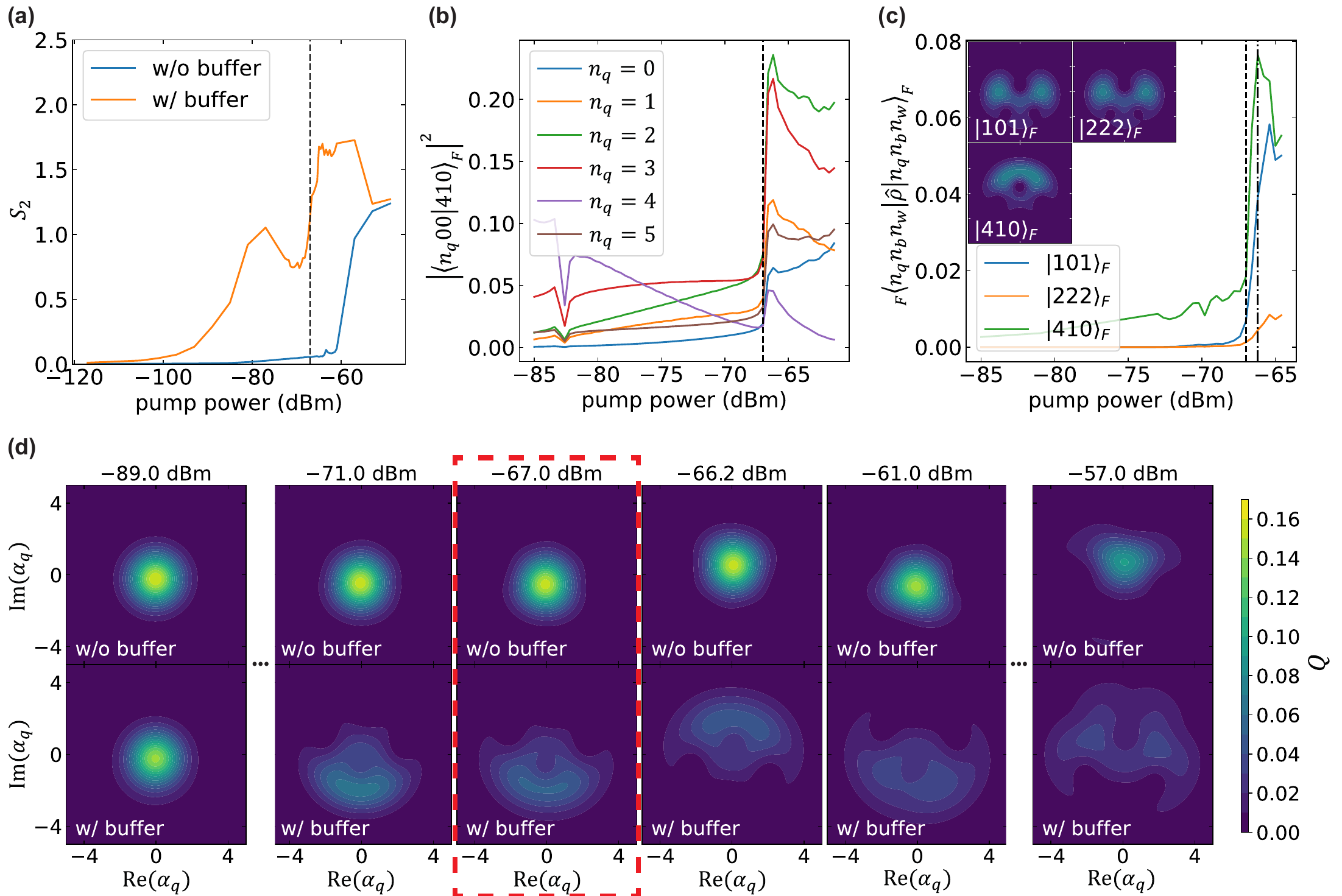}
	\caption{R{\'e}nyi entropy, Floquet simulation, and Husimi Q distribution using the Hamiltonian in Eq.\,\eqref{eq:Ham4}. We choose $\varepsilon_b=0$ for the case "without buffer drive". The vertical black dashed line marks the CPP at $P_c=-67$\,dBm. (a) R{\'e}nyi entropy, $S_2$, obtained from the simulated reduced transmon density matrix without (blue) and with (orange) the buffer drive as a function of the pump power. At the CPP, $P_c=-67$\,dBm, the R{\'e}nyi entropy remains rather smooth in the absence of buffer photons, while its sudden increase is observed for finite buffer signal powers. (b) The overlap between the transmon state $\ket{n_q00}$ and the Floquet mode $\ket{410}_F$ as a function of pump power. At the CPP, we find a drastic increase of this overlap for transmon states characterized  by different $n_q$. (c) Probability of finding specific Floquet modes in the simulated density matrix as a function of the pump power. A sudden increase of the probabilities can be observed. Insets show Husimi Q distributions of the Floquet states $\ket{101}_F$, $\ket{222}_F$ and $\ket{410}_F$ at $-66.2$\,dBm (vertical dash-dotted line). (d) Simulation of the transmon Husimi Q function in the absence (top) and presence (bottom) of the buffer signal for various pump powers. In the absence of the buffer signal, minimal variation is observed around the CPP. Conversely, in the presence of the buffer signal, the Husimi Q function spreads across the phase space, progressively converging to a double-peak distribution for higher pump powers. The red dashed box highlights the transmon Husimi Q function at the CPP.}
	\label{fig:Qfunc}
\end{figure*}

In this section, we present a systematic study of the buffer photon conversion into the transmon-waste excitations as a function of the pump power. Our measurements reveal a strong increase in the transmon population above a certain critical pump power (CPP) of $P_{c}=-67$\,dBm,  as illustrated in Fig.\,\ref{fig:nq_Pp}\,(c,d). To maximize the qubit population for each pump power, we adjust our pump frequencies accordingly. We further note that the phase shift of the transmon response $\Delta \varphi$ is directly related to the transmon population based on the dispersive readout technique \cite{Gambetta2008, Krantz2019}.

For the pump powers below the CPP, we observe a monotonic increase in the transmon population with increasing the steady-state buffer photon number. This result is in agreement with theoretical predictions, which will be discussed in the next section. Conversely, for pump powers above the CPP, we again observe a monotonic increase in transmon population, but with a discontinuity in the transmon response at the CPP (see Fig.\,\ref{fig:nq_Pp}\,(c)). In an independent measurement using a two-tone spectroscopy approach, we also find a step-like change of the buffer and waste resonance frequencies at a similar pump power, where the pump power is varied at the fixed pump frequency (see Fig.\,\ref{fig:nq_Pp}\,(e)). This observation is consistent with previous studies \cite{Reed2010,Lescanne2019} and numerical simulations \cite{Elliott2016,Verney2019,Shillito2022}, and suggests that here the transmon is entering the ionization region, where it eventually "escapes" from the Josephson potential.

We explore the system dynamics numerically by utilizing the Hamiltonian presented in Eq.\,\eqref{eq:Ham4}. By relying on experimentally determined parameters, we solve the Lindblad master equation given in Eq.\,\eqref{eq:masterOrig} and subsequently validate the proposed Hamiltonian by comparing our results with measurements shown in Fig.\,\ref{fig:nq_Pp}\,(d). During the simulation, the Hilbert space dimensions of the transmon, waste, and buffer modes are chosen to be 9, 3, and 3, respectively. The eigenenergy calculations of the transmon with the experimental parameters show that the first eight transmon states are confined states, while the ninth lies above the Josephson junction potential corresponding to the first ionized state.

To interpret the dynamics of our system, we utilize the R{\'e}nyi entropy, Floquet theory, and Husimi Q function, as shown in Fig.\,\ref{fig:Qfunc}. We begin by considering the R{\'e}nyi entropy. This quantity can offer valuable insights into the phase transition of our tripartite system around the critical point \cite{Renyi1960, Jin2004,Mavrogordatos2016, Qolibikloo2019}. Mathematically, the R{\'e}nyi entropy is defined as
\begin{equation}
S_\alpha = -\frac{1}{1-\alpha}\log_2 \text{Tr}\left( \hat{\rho}_t^\alpha \right),
\end{equation}
with the so-called order parameter $\alpha$. The reduced density matrix of the transmon, $\hat{\rho}_t$, is obtained by tracing out both the waste and buffer modes, i.e., $\hat{\rho}_t = \text{Tr}_{w,b}\left(\hat{\rho}\right)$. In this study, we choose $\alpha=2$, as this entropy serves as a key metric for quantifying the purity, and thereby, the "classicality" of the associated density matrix. Intuitively, a system with higher entropy is more mixed or less pure, and consequently, is more classical in nature. This entropy allows us to observe the quantum-to-classical phase transition. As illustrated in Fig.\,\ref{fig:Qfunc}\,(a), a distinct jump in entropy is evident at the CPP, a feature that is absent without the buffer drive. This marked change offers a clear indication of the transition of the system from a predominantly quantum state to a more classical one, thus, supporting the observed quantum-to-classical phase transition \cite{Fink2018,Verney2019,Lescanne2019, Shillito2022, Cohen2023}.

Furthermore, we apply the Floquet theory, which is particularly suited for examining strongly-driven systems, making it suitable for our quantum-to-classical phase transition studies \cite{Utermann1994,Cohen2023}. We analyze the overlap between the transmon and Floquet states as a function of the pump power. The Floquet modes are designated as $\ket{n_qn_bn_w}_F$, aligning with $\ket{n_qn_bn_w}$ at minimal pump levels, which are the eigenstates of $\hham_0$. As illustrated in Fig.\,\ref{fig:Qfunc}\,(b), a distinct transition in the overlap of the Floquet state, $\ket{410}_F$, is evident at the CPP, $P_c=-67$\,dBm. Comparable trends are noted for Floquet modes $\ket{101}_F$ and $\ket{222}_F$ (not shown). We further investigate the dynamical properties of probabilities corresponding to these Floquet modes in the resulting density matrix, deduced from the Lindblad master simulation. In line with the previous findings, a sharp increase in these probabilities is observed at the CPP, as shown in Fig.\,\ref{fig:Qfunc}\,(c). Henceforth, after the CPP, these Floquet states, $\ket{101}_F$, $\ket{222}_F$, and $\ket{410}_F$, can be viewed as predominant states characterizing the system for the increased pump powers. The insets of Fig.\,\ref{fig:Qfunc}\,(c) present the Husimi Q distributions of these Floquet states at $-66.2$\,dBm, slightly above the CPP. They reveal extensive delocalization, where $\ket{101}_F$ and $\ket{222}_F$ states are further characterized by the double-peak distributions. Here, the transmon Husimi Q function is defined as
\begin{equation*}    Q=\bra{\alpha_q}\hat{\rho}_t\ket{\alpha_q}/\pi,
\end{equation*}
where $\ket{\alpha_q}$ represents the coherent state with the complex transmon field amplitude $\alpha_q$.

Lastly, we aim to affirm that this bimodal nature observed in the Husimi Q function of the Floquet modes distinctly manifests in the phase space of the transmon after the CPP. In Fig.\,\ref{fig:Qfunc}\,(d), we show the simulation results of the Husimi Q representation of the transmon in scenarios with and without the buffer signal. We employ the double peak of the Husimi Q distribution in the phase space as an indicator of coexisting states. This characteristic bimodal structure of the Q function is a recognized feature of dissipative quantum phase transitions, typically observed in quantum Duffing oscillators \cite{Hines2005, Carmichael2015, Elliott2016, Mavrogordatos2016, Chen2023}. Thus, the identification of such a structure within our system suggests the occurrence of the first-order dissipative phase transition in the transmon-resonator system. As shown in the top panel of Fig.\,\ref{fig:Qfunc}\,(d), the Gaussian-like distribution of the Husimi Q function remains largely unchanged until around $P_p=-61$\,dBm, if the buffer drive is switched off. In other words, the delocalization does not appear even after $P_c=-67$\,dBm. This localized distribution complies with the transmon population, predominantly occupying the ground state (see Fig.\,\ref{fig:nq_Pp}\,(d)). Conversely, with the buffer drive switched on, the system undergoes a delocalization in the phase space around the CPP, progressively converging to a double-peak formation at stronger pump powers, as illustrated in the bottom panel of Fig.\,\ref{fig:Qfunc}\,(d) at $-57$\,dBm. This observation, along with prior simulation results, reveals that the delocalization in phase space is influenced by the emergence of double-peak distributions. These distributions are characterized by the Floquet modes, $\ket{101}_F$, $\ket{222}_F$, and $\ket{410}_F$. Importantly, the manifestation of this bimodal structure is not a gradual transition but occurs abruptly at the CPP. This phenomenon is consistent with the sudden increase in the transmon population observed experimentally, as shown in Fig.\,\ref{fig:nq_Pp}\,(d).

\section{single-photon detection performance}
\label{sec:spd_performance}
\begin{figure}[bt]
	\centering
	\includegraphics[width=1.\columnwidth]{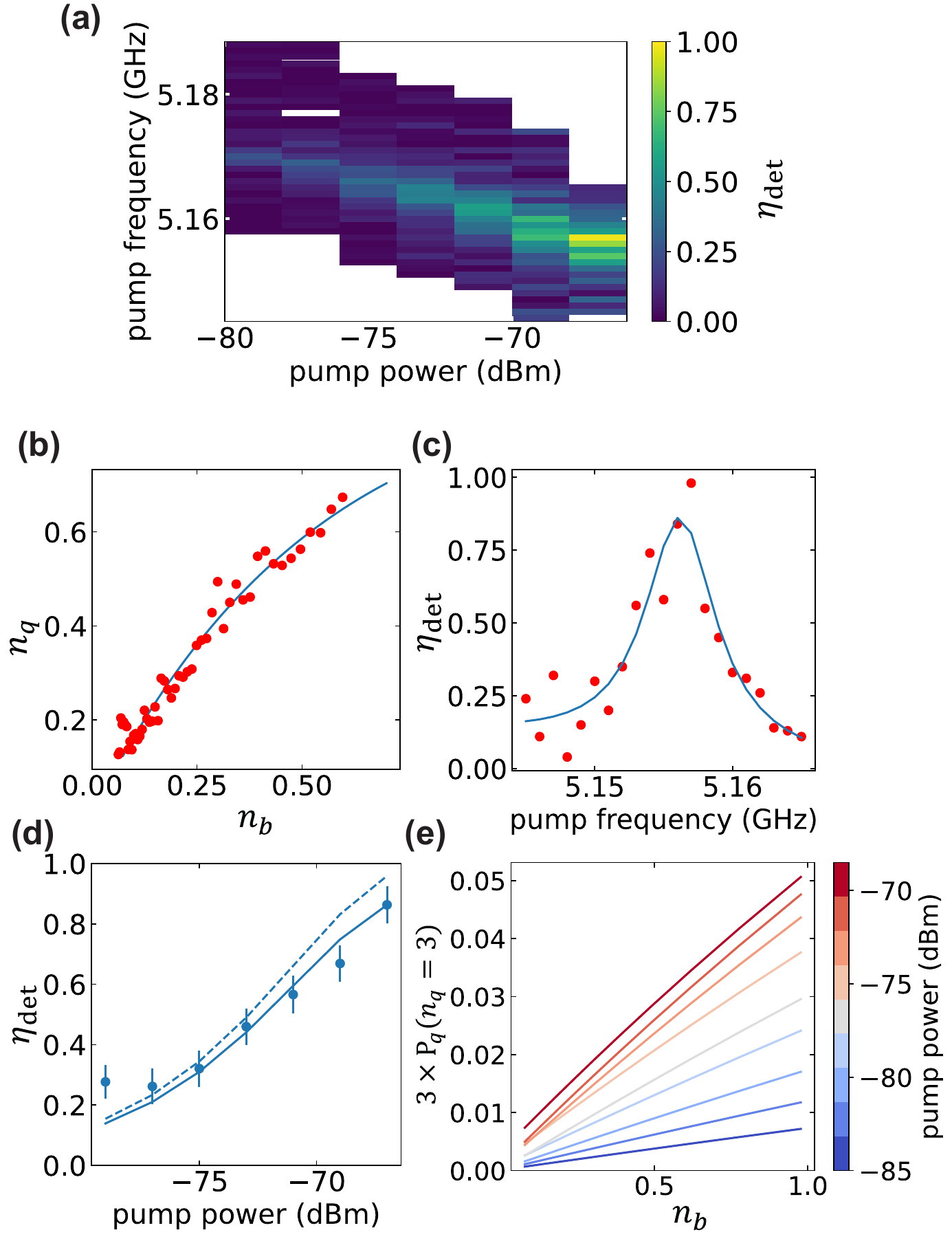}
	\caption{(a) Detection efficiency as a function of the pump frequency and power. (b) Qubit population as a function of the steady state buffer photon number $n_b$. The red dots are the measured data points, while the blue line is fitted by numerically solving the quantum Lindblad master equation with $t_b=0.55$\,$\mu$s (see Eq.\,\eqref{eq:Ham4}). (c) Lorentzian fit of $\eta_{\rm det}$ at $P_p=-67$\,dBm and $\omega_p/2\pi=5.156$\,GHz. (d) Single-photon detection efficiency $\eta_{\rm det}$ as a function of the pump power. The blue dots are the optimal detection efficiency extracted from the Lorentzian fit (blue line in (c)). The solid line is the fit using Eq.\,\eqref{sc_mw_circuit:eq:eta_c} and Eq.\,\eqref{sc_mw_circuit:eq:etadet}. The dashed line shows the case for $n^*=1$. (e) The qubit population of the state $\ket{n_q=3}$ as a function of $n_b$ for various pump powers used in the simulation. The population is calculated as $\bra{3}\hat{a}^\dagger _q \hat{a}_q\hat{\rho}_t\ket{3}=3\times \bra{3}\hat{\rho}_t\ket{3}=3\times \text{P}_q(n_q=3)$, where the probability being in the state $\ket{n_q=3}$ is denoted as $\text{P}_q(n_q=3)=\bra{3}\hat{\rho}_t\ket{3}$.}
	\label{fig:spd_eff}
\end{figure}

We further investigate our device as a single-photon detector. An important figure of merit of such a device is its detection efficiency. We extract corresponding efficiencies for various pump frequencies and powers up to the CPP at various calibrated buffer signal powers (see Fig.\,\ref{fig:spd_eff}). The performance of the conversion process is characterized by a separate conversion efficiency defined as \cite{Lescanne2020}
\begin{equation}
	\eta_c = 4 \frac{\kappa_{\rm nl} \kappa_b}{\left(\kappa_{\rm nl} + \kappa_b\right)^2},
    \label{sc_mw_circuit:eq:eta_c}
\end{equation}
with $\kappa_{\rm nl}=4 \left|g_4 \xi_p\right|^2/\kappa_w$. Its relationship with the qubit population $n_q = \left\langle \hat{\sigma}^\dagger\hat{\sigma} \right\rangle $ is
\begin{equation}
	n_q = n^*\left(1 - \exp\left(\eta_c \left| b_{\rm in}\right|^2 t_b \right)\right),
	\label{eq:nq_eta}
\end{equation}
where $\left| b_{\rm in}\right|^2$ is the buffer photon flux and $t_b$ is the buffer pulse length. Here, $n^*\leq 1$ is the saturated qubit population, whence the detection efficiency is defined as
\begin{equation}
  \eta_{\det} := n^*\eta_c.
 \label{sc_mw_circuit:eq:etadet}
\end{equation}
In actual experiments, reduced detection efficiencies, $\eta_{\det} < 1$, are attributed to factors such as a finite energy relaxation rate and finite thermal waste photon numbers, which stem from heating due to the elevated pump powers and microwave noise from the input lines \cite{Corcoles2011, Geerlings2013, Jin2015}. The detection efficiencies $\eta_{\det}$ are extracted using the Lindblad master equation simulations of the Hamiltonian given in Eq.\,\eqref{eq:Ham4}: We simulate the qubit population $n_q$ for various values of $g_4 \xi_p$, $\left|b_{\rm in} \right|^2$, and thermal bath temperatures at a fixed buffer pulse length $t_b$. We finally choose $\eta_{\rm det}$, such that the simulated $n_q(n_b)$ fits all experimentally measured curves (see Fig.\,\ref{fig:spd_eff}\,(b)). For these simulations, the qubit driving term is neglected, i.e., $\varepsilon_q = 0$. Additionally, the detuning parameter $\Delta_{qwbp}$ is set to zero, and the transmon Hilbert space dimension is set to $N_t=3$, i.e., we take the second excited state into account. These fits are conducted for every pump power and frequency, as shown in Fig.\,\ref{fig:spd_eff}\,(a). Figure\,\ref{fig:spd_eff}\,(c) shows a Lorentzian fit to obtain the optimal detection efficiencies at each pump power. Finally, we utilize the pump power dependence of $\eta_c$, such that $n^*$ and $\eta_c$ are determined by fitting the optimal detection efficiencies as a function of the pump power using Eq.\,\eqref{sc_mw_circuit:eq:eta_c} and Eq.\,\eqref{sc_mw_circuit:eq:etadet}, as demonstrated in Fig.\,\ref{fig:spd_eff}\,(d). In our studies, the maximum conversion and detection efficiencies are $\eta_c = (96\pm8)$\% and $\eta_{\rm det} = (86\pm6)$\%, respectively. In other words, we have a detection infidelity of $1-\eta_{\det} = (14\pm6)\%$. As derived in App.\,\ref{sec:sat_qubit}, by taking into account the qubit energy relaxation rate, the saturated qubit population can be determined as 
\begin{equation}
	n_{q}^* = \frac{\eta_c \left|b_{\rm in} \right|^2 }{\eta_c \left|b_{\rm in} \right|^2 + \gamma_q}.
\end{equation}
In our case, for a buffer photon flux of $\left|b_{\rm in} \right|^2 = 1.22$\,MHz corresponding to the steady-state photon number of $n_b=0.25$ we obtain $n_{q}^* = 0.97$.  Notably, although the buffer photon flux significantly surpasses the decoherence rate observed in our measurements, we can reach a detection infidelity of only a few per cent. Inserting this value into Eq.\,\eqref{sc_mw_circuit:eq:etadet}, we find the detection infidelity of 7\%.

\begin{figure*}[bt]
	\centering
	\includegraphics[width=2 \columnwidth]{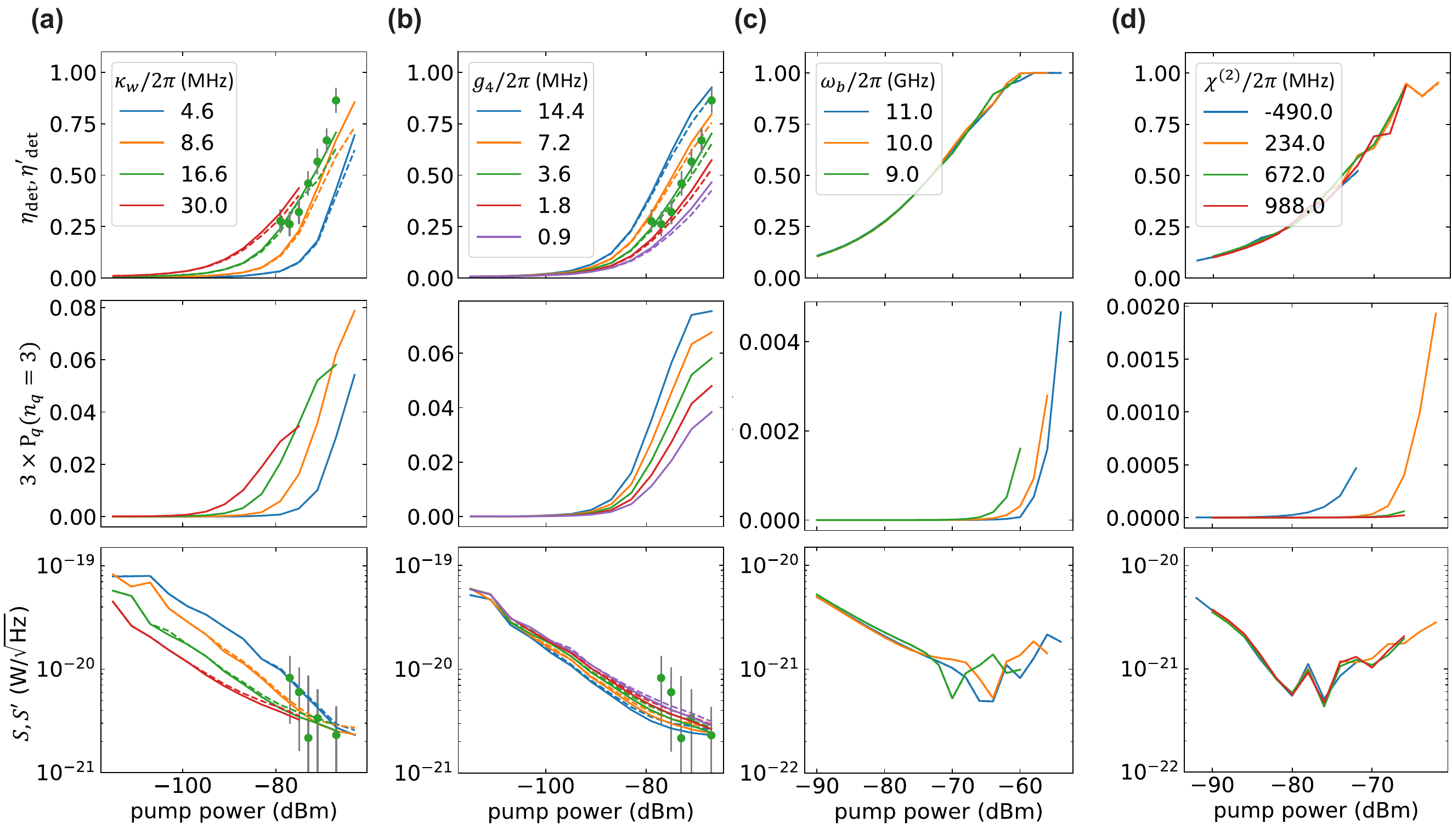}
	\caption{Detection efficiency, qubit population contribution of the state $\ket{n_q=3}$, and sensitivity as a function of the pump power for different system parameters. TI-uncorrected efficiencies $\eta_{\det}$ and sensitivities $S$ are represented by solid lines, while TI-corrected efficiencies $\eta_{\det}'$ and sensitivities $S'$ are represented by dashed lines. The qubit population of $\ket{n_q=3}$ is calculated as $\bra{3}\hat{a}^\dagger _q \hat{a}_q\hat{\rho}_t\ket{3}=3\times \text{P}_q(n_q=3)$, where the probability being in the state $\ket{n_q=3}$ is denoted as $\text{P}_q(n_q=3)$. (a) The impact of various dissipation rates of the waste mode $\kappa_w/2\pi=4.6, 8.6, 16.6, 30$\,MHz (blue, orange, green, red line). The green dots are the experimentally extracted values for $\kappa_w/2\pi=16.6$\,MHz. (b) The impact of various four-wave interaction strengths $g_4/2\pi=0.9,1.8,3.6,7.2,14.4$\,MHz (purple, red, green, orange, blue line). The green dots represent the experimentally extracted values for $g_4/2\pi=3.6$\,MHz. (c) The impact of various buffer frequencies $\omega_b/2\pi=9,10,11$\,GHz (green, orange, blue line), and (d) anharmonicities $\chi^{(2)}/2\pi=-490,234,672,988$\,MHz (blue, orange, green, red line). For the simulation with the positive anharmonicities, we consider typical flux qubit values as a reference. (a) and (b) are simulated at $T_{\rm eff}=98$\,mK, while (c) and (d) are obtained in the zero temperature limit, $T=0$\,K.}
	\label{fig:det_eff_sens_opt}
\end{figure*}

In the case of a non-zero waste photon number, the combined quantum system can partially reverse the four-wave mixing process. Given this scenario, the saturated qubit population is quantified as
\begin{equation}
	n_{w}^* = \frac{1-6n_{\text{th},w}}{1-4n_{\text{th},w}},
	\label{sc_mw_circuit:eq:SPDQubitThermal}
\end{equation}
under the condition $n_{\text{th},w}\ll1$ (see App.\,\ref{sec:sat_qubit} for more details).We employ the dark count probability of the qubit to estimate $n_{\text{th},w}$. To achieve this, we use the measurement data taken for $n_b=0$, as shown  in Fig.\,\ref{fig:nq_Pp}\,(d), and find the dark count probability of $(5.4\pm8.3)\%$, corresponding to the effective temperature of $T_{\rm eff}=98$\,mK. Using the formula $n_{\text{th},w}=(\exp(\hbar \omega_w/k_{\rm B}T_{\rm eff})-1)^{-1}$, we obtain $n_{ w}^*=0.94$. This result leads to the detection infidelity of 10\%. Adding up both infidelities, we obtain a total infidelity of 17\%, which aligns well with the measured detection infidelity of 14\%.

The discussion so far has omitted effects of the TI process. To address this, we consider the impact of the higher qubit level excitations, $\ket{n_q\geq3}$, which may additionally lead to an overestimation of the detection efficiency. For that purpose, we extract the occupation probability of $\ket{n_q\geq3}$ as a function of $n_b$ at the various pump powers, and find that the occupation probabilities of the states $\ket{n_q\geq4}$ are negligibly small. In Fig.\,\ref{fig:spd_eff}\,(e), the buffer photon number and pump power dependence of the state $\ket{n_q=3}$ can be clearly observed. Crucially, the dependence on the buffer photon number leads to an overestimation of the detection efficiency. This effect arises because the occupation of this state additionally contributes to the qubit population. In order to estimate the TI influence on the detection efficiency, we remove the contribution of higher excited states, i.e. $\sum_{n_q\leq2} \bra{n_q}\hat{a}^\dagger _q \hat{a}_q\hat{\rho}_t\ket{n_q}$, and extract its efficiency with the same method used in the experiments. Upon this adjustment, we observe a reduction in the detection efficiency by 10\% with the optimal parameters. We assume that the correction applied to the simulated data is also applicable to the experimental results. Under this assumption, the original detection efficiency of $\eta_{\rm det}=86$\% reduces to the TI-corrected detection efficiency of $\eta_{\rm det}'=78$\%.

Next, we conduct an extensive analysis to find out how to enhance the detection efficiency and sensitivity of our SPD. Furthermore, we also discuss how to suppress the influence of the TI process. Referring to Fig.\,\ref{fig:det_eff_sens_opt}, we present the TI-uncorrected and TI-corrected detection efficiencies and sensitivities across various parameters. The sensitivity is defined as \cite{Balembois2023}
\begin{align}
    S=& \frac{\hbar \omega_b \sqrt{r_{\rm dc}}}{\eta_{\rm det}}, &S'=& \frac{\hbar \omega_b \sqrt{r_{\rm dc}}}{\eta_{\rm det}'},
    \label{sc_mw_circuit:eq:sensitivity}
\end{align}
with the dark count rate $r_{\rm dc}$. We denote $S$ as the TI-uncorrected sensitivity, and $S'$ as the TI-corrected one. We observe an early rise of the detection efficiency $\eta_{\rm det}$ ($\eta_{\rm det}'$) for the higher dissipation rate of the waste mode $\kappa_w$, as demonstrated in Fig.\,\ref{fig:det_eff_sens_opt}\,(a). However, the increase of $\kappa_w$ also leads to an enhancement of the Purcell decay rate of the qubit. Thus, $\varepsilon_q$ increases for the same pump power accelerating the onset of the TI, which sets the upper bound of the conversion process. Consequently, despite the early rise in detection efficiency, $\eta_{\rm det}$ ($\eta_{\rm det}'$) fails to achieve higher values for larger $\kappa_w$. The early increase of the probability $\text{P}_q(n_q=3)$ supports the argument of this early onset of the TI. Therefore, the optimal detection efficiency, $\eta_{\rm det}$ ($\eta_{\rm det}'$), along with the sensitivity $S$ ($S'$), is more favorably achieved at lower $\kappa_w$. As shown in Fig.\,\ref{fig:det_eff_sens_opt}\,(b), an enhancement in the coupling strength $g_4$ improves both $\eta_{\rm det}$ ($\eta_{\rm det}'$) and $S$ ($S'$). Since the CPP does not change under $g_4$ variation, we can reach higher $\eta_{\rm det}$ ($\eta_{\rm det}'$) before reaching the CPP. Additionally, when we tune the buffer frequency $\omega_b$ to higher frequencies, Fig.\,\ref{fig:det_eff_sens_opt}\,(c) predicts that $\text{P}_q(n_q=3)$ decreases. This is attributed to a large detuning between pump and qubit frequencies, which reduces the TI impact. In other words, $\eta_{\rm det}$ converges to $\eta_{\rm det}'$. Finally, we investigate the influence of the TI for different anharmonicities $\chi^{(2)}$. We observe that large $\left|\chi^{(2)}\right|$ values suppress the TI process, as expected from the previous studies \cite{Gambetta2011, Chen2016, Yan2016}. Accordingly, $\eta_{\rm det}$ asymptotically approaches $\eta_{\rm det}'$. Notably, the configuration of the buffer and waste frequencies should be chosen considering a particular value of $\chi^{(2)}$. In general, the pump frequency can match a frequency of a certain multi-photon process at a certain pump power due to the power-dependent  frequency shift of the qubit frequency. In the regime of negative anharmonicity, this situation can occur if $\omega_w<\omega_b$, such that $\omega_p,\omega_{0k}/k<\omega_q'$, where $\omega_{0k}/k$ is the multi-photon transition frequency. If the pump frequency is close to such a transition frequency, the dark count rate increases, which eventually degrades the SPD performance. For example for $\chi^{(2)}/2\pi=-490$\,MHz, the SPD can reach $\eta_{\rm det}\approx 0.5$, until the pump collides with the two-photon process in the spectral domain. In simulations with positive anharmonicity values, we use a flux qubit configuration \cite{Yan2016}, and achieve detection efficiencies up to $95$\%. Importantly, for the flux qubit, multi-photon processes become relevant at frequencies exceeding the qubit frequency, and given that $\omega_p<\omega_q'$ for $\omega_w<\omega_b$, such processes are effectively far-detuned, resulting in a reduction of dark counts (Fig.\,\ref{fig:det_eff_sens_opt}\,(d)).

Lastly, we compare the sensitivity between the cases with $T_{\rm eff}=98$\,mK (Fig.\,\ref{fig:det_eff_sens_opt}\,(a,b)) and $T=0$\,K (Fig.\,\ref{fig:det_eff_sens_opt}\,(c,d)). Although the sensitivity for an ideal environment ($T=0$\,K) generally improves, it does not reach the present state of the art value of $S=10^{-22}\ \rm W/\sqrt{Hz}$ \cite{Balembois2023}. This is attributed to the filtering effect of our cavity, which requires stronger pump powers to reach the same detection efficiency, and hence, results in a higher dark count probability due to the pump heating. In our case, the dark count rate gives $26\pm4$\,kHz at the pump power of $P_p=-67$\,dBm, at which we measure the maximum $\eta_{\rm det}$ ($\eta_{\rm det}'$). Despite of the cavity filtering effect, the dark count rate reported here is comparable to the values reported in the previous studies \cite{Inomata2016, Kono2018, Besse2018}.

\section{Conclusion}
\label{sec:conclusion}
In this work, we have presented an experimental and numerical study of the TI onset and the SPD performance of a transmon qubit coupled to a multi-mode 3D cavity. We have investigated the dependence of the transmon population on the buffer photon number, $n_b$, and pump power, and observed that the transmon is highly-sensitive to $n_b$ within the region close to a critical pump powers (CPP). Through a comprehensive analysis employing the R{\'e}nyi entropy, Floquet theory, and Husimi Q function, we have elaborated an occurrence of a quantum-to-classical phase transition around the CPP. Our numerical simulations align well with experimental observations, emphasizing the accuracy of our chosen methodologies.

We have also investigated the device as a single-photon detector, and extracted its detection efficiencies for various pump frequencies and powers up to the CPP. Our measurements show that a maximum TI-uncorrected detection efficiency of 86\% can be achieved, while we estimate the TI-corrected detection efficiency to be 78\%. The conversion efficiency $\eta_c$ between the buffer and the qubit-waste subsystems is limited by the TI, while the reduction of the saturated qubit population $n^*$ can be attributed to a finite thermal waste photon number and the qubit energy relaxation rate. Finally, increasing the frequency detuning between the pump and qubit frequencies and the qubit anharmonicity is expected to strongly suppress the influence of the TI process and leads to a distinct improvement of the TI-corrected detection efficiency and sensitivity. The spectral positioning of the buffer and waste frequencies have to be determined based on the sign of the qubit anharmonicity to ensure that the pump and multi-photon processes do not overlap.

In conclusion, our systematic studies reveal the specific advantages inherent to the onset of TI, suggesting its prospective extension to high-efficiency microwave SPDs compatible with 3D cavity architectures. Moreover, our study emphasizes the necessity for a careful analysis of the efficiency extraction, particularly given the impact of the TI process. These insights bear significant implications for the development of parametric device applications, which are essential in advancing quantum information processing and communication with superconducting circuits.

\section*{Acknowledgments}
We thank Joan Agusti, Yvan Briard, Louis Garbe and Peter Rabl for insightful discussions. We acknowledge support by the German Research Foundation via Germany’s Excellence Strategy (EXC2111-390814868), the German Federal Ministry of Education and Research via the project QUARATE (Grant No.13N15380). This research is part of the Munich Quantum Valley, which is supported by the Bavarian state government with funds from the Hightech Agenda Bayern Plus.

\appendix
\section{Experimental techniques}
\label{sec:ExpTech}

\begin{figure}[bt]
	\centering
	\includegraphics[width=0.75\columnwidth]{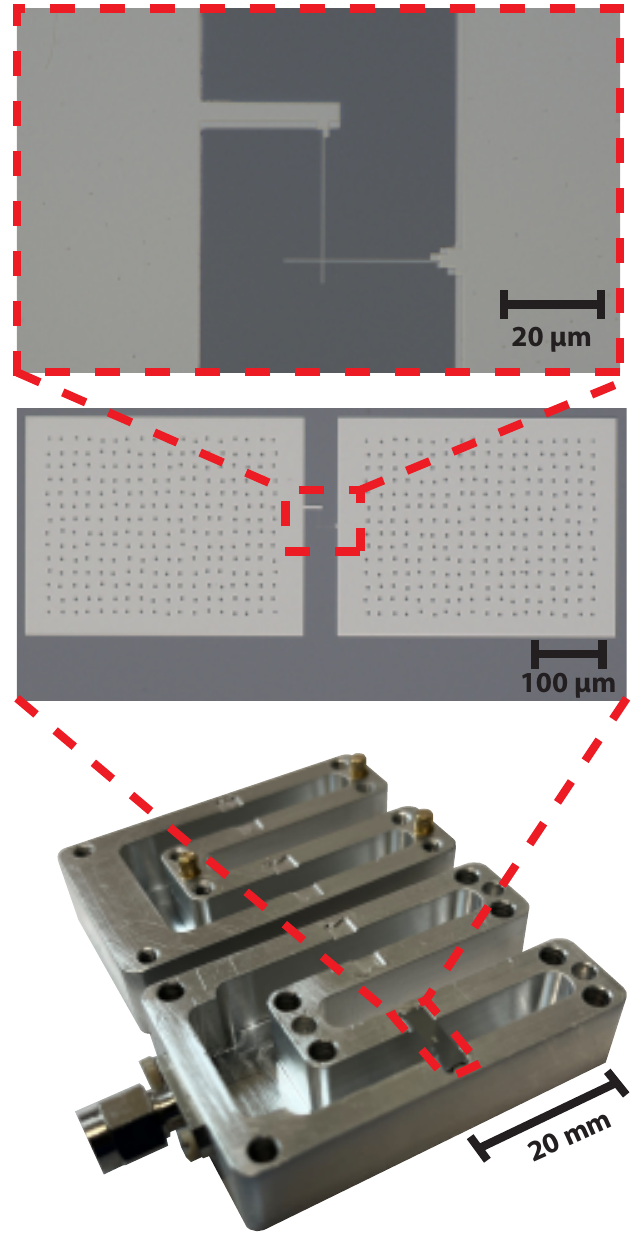}
	\caption{Picture of the device. (top) Optical image of the Josephson junction. (center) Optical image of the transmon qubit with its antenna pads. (bottom) Photograph of two halves of the Al horseshoe cavity with the transmon qubit chip inside. }
\label{fig:spd_sample}
\end{figure}

\subsection{Microwave single-photon detector sample}
As outlined in Sec.\,\ref{sec:model}, our single-photon detection device is comprised of two primary components: a horseshoe cavity and a transmon qubit chip. The cavity, constructed from aluminum (Al) with a high purity level of $99.99$\%, has been fabricated by the workshop at the Walther-Mei{\ss}ner-Institut. To address the relatively low thermal conductance of superconducting aluminum, we mount a gold-plated copper plate, 1.5\,mm in thickness, onto the cavity. This plate significantly enhances the thermal conductance of the system, ensuring more efficient heat removal. 

The transmon qubit, central to our detection mechanism, is fabricated on a silicon chip with spatial dimension of $3.5$\,mm by $10$\,mm. For the fabrication of the transmon qubit, a standard lift-off technique is used. We evaporate aluminum through a PMMA/CSAR resist mask written in an electron-beam lithography step. The antenna of the qubit has a size of $890$\,$\mu$m by $330$\,$\mu$m, ensuring a sufficient dipole interaction strength with the cavity. Additionally, the Josephson junctions have an area of $240 \times  240\,\text{nm}^2$.

A visual representation of the assembly, showing the integration of the open horseshoe cavity with the embedded transmon qubit chip, can be found in Fig.\,\ref{fig:spd_sample}.

\begin{figure*}[tb]
	\centering
	\includegraphics[width=1.75\columnwidth]{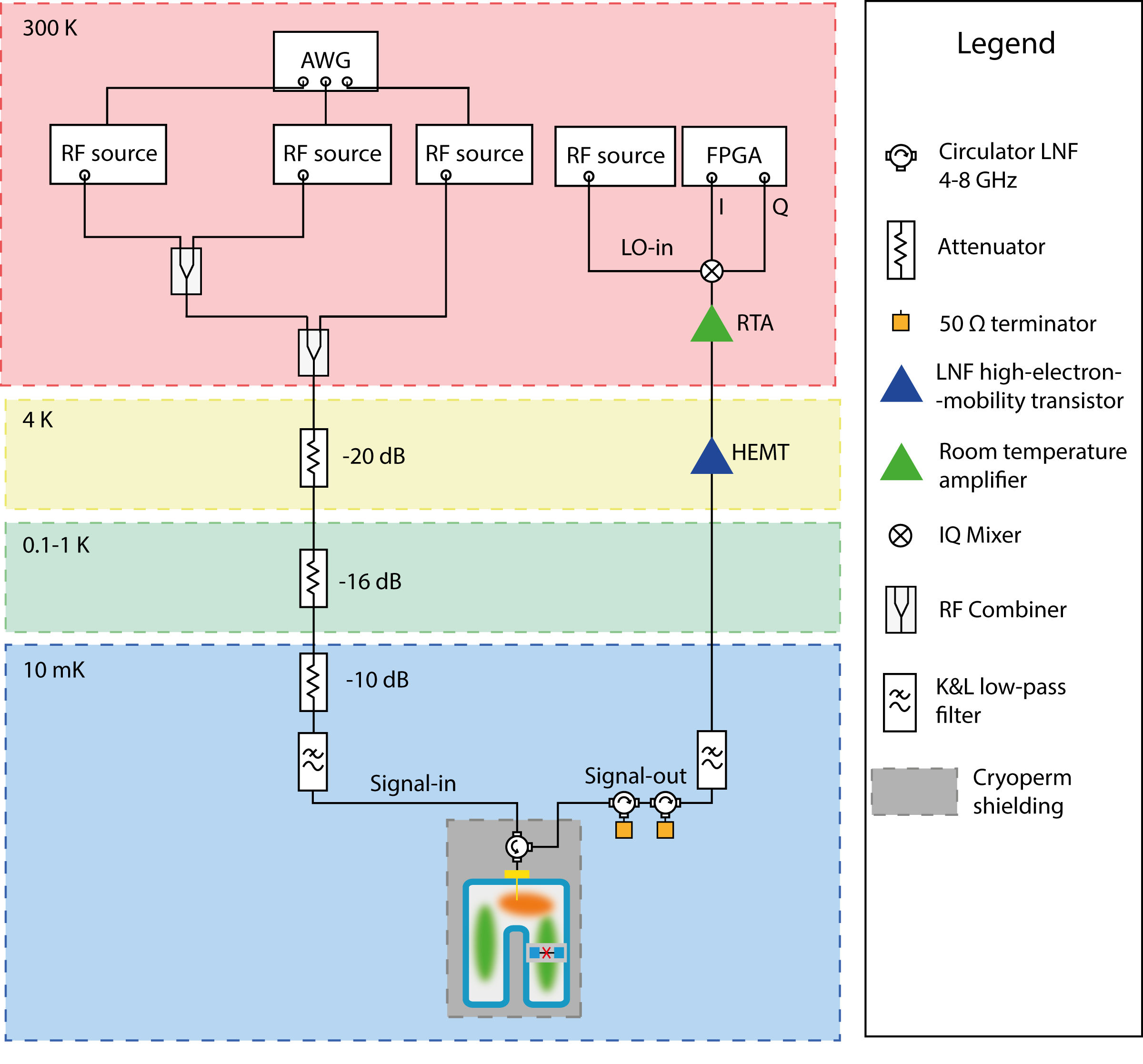}
	\caption{Schematic drawing of dry dilution refrigerator embedding Al horseshoe cavity with the transmon qubit chip inside.}\label{fig:Circuit_cooldown_SPD_02}
\end{figure*}

\subsection{Cryogenic setup}
The cryogenic arrangement utilized for our experimental measurements is illustrated in Fig.\,\ref{fig:Circuit_cooldown_SPD_02}.

For the generation of arbitrary waveforms, we employ the HDAWG from Zurich Instruments. Additionally, the setup incorporates three microwave sources (R\&S SGS 100A). To integrate these three pulsed drives, we utilize two power combiners. Initially, the pump and buffer pulses are merged, followed by their subsequent combination with the readout pulse. This combined signal is then directed to the cryostat, facilitating both the frequency conversion process and the qubit readout (see also Fig.\,\ref{fig:SPD_principle} in the main text).

The combined signals travel through a series of attenuators mounted at various cooling stages within the cryostat, followed by a low-pass filter to suppress high-frequency noise. Upon reaching the mixing chamber stage, the signal is delivered to the cavity-qubit system. The Al horseshoe cavity, housing the transmon qubit chip, is shielded from external magnetic fields by a cryoperm shield. The single-port design of the cavity is coupled with a circulator to separate the output signal from the incoming signal.

The output path includes a low-pass filter, followed by passage through two isolators. Amplification is achieved via a HEMT amplifier located at the 4\,K stage, and subsequently, through a room-temperature amplifier.

\section{Saturated qubit population}
\label{sec:sat_qubit}
In this section, we derive the influence of the qubit decoherence and residual waste photon number on the saturated qubit population. For that purpose, we briefly review the derivation of the nonlinear decay by tracing out the waste modes with adiabatic elimination \cite{Lescanne2020}. After canceling out the ac-Stark terms, neglecting the driving terms and restricting the Hilbert space of the qubit mode to a two-level system ($\hat{a}_q\rightarrow \hat{\sigma}$) in Eq.\,\eqref{eq:Ham4}, we arrive at
\begin{align*}
	\hham''_4=& \hham_w + \hham_{qb}\\
	\hham_w/\hbar = & \left(\Delta_{w} - \chi_{qw} \hat{\sigma}^\dagger \hat{\sigma} \right)\hat{a}_w^\dagger \hat{a}_w + g_{4} \xi_p \hat{a}_b \hat{a}_w^\dagger \hat{\sigma}^\dagger + \rm h.c.\\
	\hham_{qb}/\hbar =& \chi_{qb}\hat{\sigma}^\dagger \hat{\sigma} {\hat{a}_{b}}^\dagger \hat{a}_{b},
\end{align*}
which dynamics is described by the Lindblad master equation
\begin{align}
	\frac{\text{d}}{\text{d}t}\hat{\rho} =&  -i \left[\hham_w,\hat{\rho}\right]  +\mathcal{L}_{qb}\left[\hat{\rho}\right] \nonumber\\
    &+ \kappa_w\left(n_{\text{th},w}+1\right)\mathcal{D}[\hat{a}_w]\hat{\rho} + \kappa_wn_{\text{th},w} \mathcal{D}[\hat{a}_w^\dagger]\hat{\rho},
	\label{eq:masterOrig}
\end{align}
where $n_{\text{th},w}=(\exp(\hbar \omega_w/k_{\rm B}T_{\rm eff})-1)^{-1}$ and
\begin{align*}
	\mathcal{L}_{qb}\left[\hat{\rho}\right] =& -i \left[\hham_{qb},\hat{\rho}\right]  + \kappa_b\mathcal{D}[\hat{a}_b]\hat{\rho}  + \gamma_q\mathcal{D}[\hat{\sigma}]\hat{\rho}.
\end{align*}

For the adiabatic elimination of the waste mode, we assume
\begin{equation*}
	\left|  \frac{g_{4}\xi_p}{\kappa_w}\right| , \left|  \frac{\chi_{jj'}}{\kappa_w}\right|\sim \delta \ll 1,
\end{equation*}
and $n_{\text{th},w}=0$, such that the waste mode is dominantly in the vacuum state due to the fast decay rate $\kappa_w$. Hence, we can reduce the Hilbert space of the waste mode to $\mathcal{H}_w=\text{span}\left( \ket{0_w},\ket{1_w}\right) $. In particular, for the density matrix $\hat{\rho}\in \mathcal{H}_q\otimes\mathcal{H}_b\otimes\mathcal{H}_w$, the reduced density matrices acting on the qubit and buffer Hilbert space have the following relations
\begin{align*}
	\bra{0_w}\hat{\rho}\ket{0_w} =&\hat{\rho}_{00}, & \bra{0_w}\hat{\rho}\ket{1_w} =& \delta\hat{\rho}_{01}, \nonumber\\
	\bra{1_w}\hat{\rho}\ket{1_w} =& \delta^2\hat{\rho}_{11}, &	\bra{0_w}\hat{\rho}\ket{2_w} =& \delta^2\hat{\rho}_{02},
\end{align*}
with $\ket{j_w}, \ket{j'_w}$ being the Fock basis of the waste mode. Note that $\hat{\rho}_{00}\in \mathcal{H}_q\otimes\mathcal{H}_b$. By projecting Eq.\,\eqref{eq:masterOrig} with $\bra{0_w}...\ket{0_w},\bra{0_w}...\ket{1_w},\bra{1_w}...\ket{1_w}$, respectively, we obtain
\begin{align}
	\frac{\text{d}}{\text{d}t}\hat{\rho}_{00}
	= & -i \delta \left[g_{4}\xi_p^* \hat{a}_b^\dagger \hat{\sigma} \hat{\rho}_{10}  -g_{4}\xi_p\hat{\rho}_{01} \hat{a}_b \hat{\sigma}^\dagger\right] + \delta^2 \kappa_w\hat{\rho}_{11} \nonumber\\
    &+\mathcal{L}_{qb}\left[\hat{\rho}_{00}\right] + \mathcal{O}\left(\delta^3\right)
	\label{eq:rho00}\\
	\delta \frac{\text{d}}{\text{d}t}\hat{\rho}_{01}
	=& i\hat{\rho}_{00}\left[g_{4}\xi_p^* \hat{a}_b^\dagger \hat{\sigma} \right]+i\delta\hat{\rho}_{01}\left[\Delta_{w} - \chi_{qw} \hat{\sigma}^\dagger \hat{\sigma}  \right] \nonumber\\
    &- \delta \frac{\kappa_w}{2}\hat{\rho}_{01} +\delta \mathcal{L}_{qb}\left[\hat{\rho}_{01}\right]  + \mathcal{O}\left(\delta^3\right)
	\label{eq:rho01}\\
	\delta^2\frac{\text{d}}{\text{d}t}\hat{\rho}_{11} 
	=& -i \delta \left[g_{4}\xi_p \hat{a}_b \hat{\sigma}^\dagger \hat{\rho}_{01}  -g_{4}\xi_p^* \hat{\rho}_{10} \hat{a}_b^\dagger \hat{\sigma}  \right] \nonumber\\
    &+i\delta^2 \left[\Delta_{bwq} - \chi_{qw} \hat{\sigma}^\dagger\hat{\sigma}, \hat{\rho}_{11}\right]\nonumber\\
	&- \delta^2 \kappa_w \hat{\rho}_{11}+\delta^2 \mathcal{L}_{qb}\left[\hat{\rho}_{11}\right]   + \mathcal{O}\left(\delta^3\right)
	\label{eq:rho11}
\end{align}
Focusing on the relevant dynamics, we find that Eqs.\,\eqref{eq:rho01} and \eqref{eq:rho11} include a damping term of order $\delta^0$, while all terms in Eq.\,\eqref{eq:rho00} are of order $\delta^2$. Hence, this allows us to treat $\hat{\rho}_{01}$ and $\hat{\rho}_{11}$ as a steady state (adiabatic approximation), which leads to
\begin{align}
	\hat{\rho}_{01}
	\approx&\hat{\rho}_{00} \frac{ig_{4}\xi_p^*/\delta}{\frac{\kappa_w}{2} -i \left(\Delta_{w} - \chi_{qw}\right)}\hat{a}_b^\dagger \hat{\sigma},
	\label{eq:rho01ss}
\end{align}
where we have used $\hat{a}_b^\dagger \hat{\sigma} \hat{\sigma}^\dagger \hat{\sigma}  = \hat{a}_b^\dagger \sigma$. As for the steady state solution for $\hat{\rho}_{11}$, we get
\begin{align}
	\hat{\rho}_{11} =& \frac{1}{\left(\frac{\kappa_w}{2}\right)^2 + \left(\Delta_{w} - \chi_{qw}\right)^2} \frac{\left| g_{4}\xi_p \right| ^2}{\delta^2} \hat{a}_b \hat{\sigma}^\dagger \hat{\rho}_{00} \hat{a}_b^\dagger \hat{\sigma}.
\end{align}
Inserting the steady state solutions into Eq.\,\eqref{eq:rho00}, we obtain
\begin{align}
	\frac{\text{d}}{\text{d}t}\hat{\rho}_{00} 
	=&\left[\frac{ i\left| g_{4}\xi_p\right| ^2\left(\Delta_{w} - \chi_{qw}\right)}{\left(\frac{\kappa_w}{2}\right)^2 + \left(\Delta_{w} - \chi_{qw}\right)^2}\hat{a}_b^\dagger \hat{a}_b \hat{\sigma} \hat{\sigma}^\dagger, \hat{\rho}_{00}\right] \nonumber\\
	&+\kappa_{\rm nl}\mathcal{D}\left[ \hat{a}_b \hat{\sigma}^\dagger\right]\hat{\rho}_{00} +\mathcal{L}_{qb}\left[\hat{\rho}_{00}\right] 
	\label{eq:rho00ss}
\end{align}
with the nonlinear decay rates $\hat{L}_{\rm nl} = \sqrt{\kappa_{\rm nl}} \hat{a}_b \hat{\sigma}^\dagger$
\begin{align}
	\kappa_{\rm nl}&\equiv \frac{\left| g_{4}\xi_p\right| ^2}{\left(\frac{\kappa_w}{2}\right)^2 + \left(\Delta_{w} - \chi_{qw}\right)^2} \kappa_w \stackrel{\Delta_w=\chi_{qw}}{=} \frac{4\left| g_{4}\xi_p\right| ^2}{\kappa_w}.
\end{align}\\

\subsection{Influence of qubit decoherence}
Here, we delve into the impact of qubit decoherence on its dynamics. Notably, it becomes evident that the qubit population never reaches unity. This is a direct implication of the fact that qubit decoherence inherently reduces the detection efficiency in single-photon detection scenarios.

The qubit decoherence term and the buffer driving term $i\varepsilon_b'\left(\hat{a}_b-\hat{a}_b^\dagger\right)$ are added to the Lindblad master equation in Eq.\,\eqref{eq:rho00ss}, such that we obtain
\begin{align}
	\frac{\text{d}}{\text{d}t}\hat{\rho}_{00} 
	= & \kappa_{\rm nl}\mathcal{D}\left[ \hat{a}_b \hat{\sigma}^\dagger\right]\hat{\rho}_{00}  +\kappa_{b}\mathcal{D}\left[  \hat{a}_b\right]\hat{\rho}_{00} \nonumber\\
    &+ \varepsilon_b'\left[\hat{a}_b-\hat{a}_b^\dagger, \hat{\rho}_{00}\right] + \gamma_q \mathcal{D}\left[  \hat{\sigma}\right]\hat{\rho}_{00}.
\end{align}
Since the time evolution of the buffer mode is dependent on the qubit state, we first consider $\hat{\rho}_g\equiv\bra{g}\hat{\rho}_{00}\ket{g}$ and $\hat{\rho}_e\equiv\bra{e}\hat{\rho}_{00}\ket{e}$, such that
\begin{subequations}
	\begin{align}
		\frac{\text{d}}{\text{d}t}\hat{\rho}_{g} 
		= & - \frac{1}{2}\kappa_{\rm nl} \left( \hat{\rho}_g  \hat{a}_b \hat{a}_b^\dagger +  \hat{a}_b \hat{a}_b^\dagger \hat{\rho}_g  \right) +\kappa_{b}\mathcal{D}\left[  \hat{a}_b\right]\hat{\rho}_{g}\nonumber\\
        & + \varepsilon_b' \left[\hat{a}_b-\hat{a}_b^\dagger, \hat{\rho}_{g}\right]  + \gamma_q \hat{\rho}_e,\\
		\frac{\text{d}}{\text{d}t}\hat{\rho}_{e} 
		= &\kappa_{\rm nl}\hat{a}_b \hat{\rho}_g \hat{a}_b^\dagger +\kappa_{b}\mathcal{D}\left[  \hat{a}_b\right]\hat{\rho}_{e} \nonumber\\
        &+ \varepsilon_b' \left[\hat{a}_b-\hat{a}_b^\dagger, \hat{\rho}_{e}\right] - \gamma_q \hat{\rho}_e.
	\end{align}
\end{subequations}
In our case, since the applied buffer tone is a coherent tone, we can safely assume that the buffer is in a coherent state $\ket{\beta}$ with the dimensionless amplitude $\beta = -2\varepsilon'_b/(\kappa_{\rm nl}+\kappa_{b})$. That is, $\hat{\rho}_{g/e} \propto \ket{\beta}\bra{\beta}$. We can now calculate the qubit ground and excited state probability by tracing them, $p_{g/e} = \text{Tr}(\hat{\rho}_{g/e} )$. This leads us to formulate the differential equation as
\begin{align}
	\frac{\text{d}}{\text{d}t}\left[ \begin{array}{c}
		p_{g}\\
		p_{e}
	\end{array}\right]  
	= & \left[ \begin{array}{cc}
		- \eta_c  \left|b_{\rm in} \right|^2 & \gamma_q \\
		\eta_c  \left|b_{\rm in} \right|^2 & -\gamma_q
	\end{array}\right] \left[ \begin{array}{c}
		p_{g}\\
		p_{e}
	\end{array}\right] ,
\end{align}
which solution gives
\begin{widetext}
\begin{align}
	\left[ \begin{array}{c}
		p_{g}(t)\\
		p_{e}(t)
	\end{array}\right]  
	= & \left[ \begin{array}{cc}
		\frac{\eta_c \left|b_{\rm in} \right|^2 }{\eta_c \left|b_{\rm in} \right|^2 + \gamma_q} e^{-\eta_c  \left|b_{\rm in} \right|^2 t - \gamma_q t} + \frac{\gamma_q }{\eta_c \left|b_{\rm in} \right|^2 + \gamma_q} & \frac{\gamma_q }{\eta_c \left|b_{\rm in} \right|^2 + \gamma_q} \left(1-e^{-\eta_c  \left|b_{\rm in} \right|^2 t - \gamma_q t} \right)  \\
		\frac{\eta_c \left|b_{\rm in} \right|^2 }{\eta_c \left|b_{\rm in} \right|^2 + \gamma_q}\left(1-e^{-\eta_c  \left|b_{\rm in} \right|^2 t - \gamma_q t} \right) & \frac{\eta_c \left|b_{\rm in} \right|^2 }{\eta_c \left|b_{\rm in} \right|^2 + \gamma_q} + \frac{\gamma_q }{\eta_c \left|b_{\rm in} \right|^2 + \gamma_q} e^{-\eta_c  \left|b_{\rm in} \right|^2 t - \gamma_q t}
	\end{array}\right] \left[ \begin{array}{c}
		p_{g}(0)\\
		p_{e}(0)
	\end{array}\right].
\end{align}
\end{widetext}
We find the excitation probability of the transmon qubit for sufficiently long time $t$ is
\begin{equation}
	p_e(t\rightarrow\infty) = \frac{\eta_c \left|b_{\rm in} \right|^2 }{\eta_c \left|b_{\rm in} \right|^2 + \gamma_q}<1.
\end{equation}
It is noteworthy that the saturation level of the qubit population shows a relative dependency on the qubit energy relaxation rate $\gamma_q$, and the "effective buffer photon conversion rate" $\eta_c \left|b_{\rm in} \right|^2$. Particularly, the impact of the energy relaxation rate becomes significant when the value of $\eta_c \left|b_{\rm in} \right|^2$ is comparably small. This observation underscores an interplay or a steady-state dynamics between the information gain from the incoming buffer photon and its subsequent loss.

\subsection{Influence of residual waste photon number}

In our prior discussions, we assumed that our quantum system experienced only vacuum noise, excluding the influence of the coherent drive. This is a standard assumption, primarily justified by the extremely low temperatures of dilution refrigerators, typically in the range of $10-50$\,mK. Nevertheless, when it comes to parametric devices, strong system pumping can induce an effective temperature increase for the quantum apparatus. This is particularly relevant in our case: the presence of a residual waste photon number can reduce detection efficiency. This is attributed to the fact that a partial reversal process ($\hat{a}^\dagger_b \hat{a}_w \hat{\sigma}$) is allowed. Given this potential complication, it is imperative to thoroughly examine how the residual waste photon number affects the qubit's temporal evolution.

We model the residual waste photon number as arising from a thermal bath. Hence, the Eqs.\,\eqref{eq:rho00}, \eqref{eq:rho01} and \eqref{eq:rho11} are modified to
\begin{align}
	\frac{\text{d}}{\text{d}t}\hat{\rho}_{00} 
	= & -i \delta \left[g_{4}\xi_p^* \hat{a}_b^\dagger \hat{\sigma} \hat{\rho}_{10}  -g_{4}\xi_p\hat{\rho}_{01} \hat{a}_b \hat{\sigma}^\dagger\right] \nonumber\\
    &+ \delta^2 \kappa_w \left(1+n_{\text{th},w} \right)\hat{\rho}_{11} - \kappa_w n_{\text{th},w} \hat{\rho}_{00}  \nonumber\\
    &+\mathcal{L}_{qb}\left[\hat{\rho}_{00}\right] + \mathcal{O}\left(\delta^3\right)\\
	\delta \frac{\text{d}}{\text{d}t}\hat{\rho}_{01}
	=& i\hat{\rho}_{00}\left[g_{4}\xi_p^* \hat{a}_b^\dagger \hat{\sigma} \right]- \delta \frac{\kappa_w \left(1+4n_{\text{th},w} \right)}{2}\hat{\rho}_{01} \nonumber\\
    &+\delta \mathcal{L}_{qb}\left[\hat{\rho}_{01}\right]  + \mathcal{O}\left(\delta^3\right)\\
	\delta^2\frac{\text{d}}{\text{d}t}\hat{\rho}_{11} 
	=& -i \delta \left[g_{4}\xi_p \hat{a}_b \hat{\sigma}^\dagger \hat{\rho}_{01}  -g_{4}\xi_p^* \hat{\rho}_{10} \hat{a}_b^\dagger \hat{\sigma}  \right] \nonumber\\
	&- \delta^2 \kappa_w \left(1+3n_{\text{th},w} \right) \hat{\rho}_{11}+ n_{\text{th},w}\hat{\rho}_{00} \nonumber\\
    &+\delta^2 \mathcal{L}_{qb}\left[\hat{\rho}_{11}\right]   + \mathcal{O}\left(\delta^3\right)
\end{align}
Here, we set $\Delta_{w} = \chi_{qw}$ for simplicity. With the same argument, we can find the steady state solutions for $\hat{\rho}_{01}$ and $\hat{\rho}_{11}$
\begin{align*}
	\hat{\rho}_{01} \approx&\hat{\rho}_{00} \frac{2 ig_{4}\xi_p^*}{\delta\kappa_w (1+4n_{\text{th},w})}\hat{a}_b^\dagger \hat{\sigma}\\
	\hat{\rho}_{11} \approx& \frac{1}{\delta^2 \kappa_w \left(1+3n_{\text{th},w}\right)}  \frac{4 \left| g_{4}\xi_p \right| ^2}{\kappa_w (1+4n_{\text{th},w})} \hat{a}_b \hat{\sigma}^\dagger \hat{\rho}_{00} \hat{a}_b^\dagger \hat{\sigma} \nonumber\\
    &+ \frac{1}{\delta^2 \kappa_w \left(1+3n_{\text{th},w}\right)} \kappa_w n_{\text{th},w} \hat{\rho}_{00}
\end{align*}
and insert them into
\begin{align}
	\frac{\text{d}}{\text{d}t}\hat{\rho}_{00}
	\approx& \kappa_{\rm nl} \left(1-4n_{\text{th},w}\right) \mathcal{D}[\hat{a}_b \hat{\sigma}^\dagger]\hat{\rho}_{00} \nonumber\\
    &- 2 \kappa_{\rm nl} n_{\text{th},w}  \hat{a}_b \hat{\sigma}^\dagger \hat{\rho}_{00} \hat{a}_b^\dagger \hat{\sigma} +\mathcal{L}_{qb}\left[\hat{\rho}_{00}\right]\nonumber\\
    & + \mathcal{O}\left(\delta^3, n_{\text{th},w}^2\right).
\end{align}
The Lindblad master equation therefore is read with the buffer driving term $i\varepsilon_b'\left(\hat{a}_b-\hat{a}_b^\dagger\right)$ as 
\begin{align}
	\frac{\text{d}}{\text{d}t}\hat{\rho}_{00} 
	= & \kappa_{\rm nl} \left(1-4n_{\text{th},w}\right)  \mathcal{D}\left[ \hat{a}_b \hat{\sigma}^\dagger\right]\hat{\rho}_{00} \nonumber\\
    &- 2 \kappa_{\rm nl} n_{\text{th},w}  \hat{a}_b \hat{\sigma}^\dagger \hat{\rho}_{00} \hat{a}_b^\dagger \hat{\sigma} +\kappa_{b}\mathcal{D}\left[  \hat{a}_b\right]\hat{\rho}_{00}\nonumber\\
    & + \varepsilon_b'\left[\hat{a}_b-\hat{a}_b^\dagger, \hat{\rho}_{00}\right].
    \label{sc_mw_circuit:eq:SPDMasterThermal}
\end{align}
After following the same procedure as outlined previously, the differential equation is formulated as
\begin{subequations}
	\begin{align}
		\frac{\text{d}}{\text{d}t}p_{g} 
		= & - \kappa_{\rm nl} \left(1-4n_{\text{th},w}\right)\left|\beta' \right|^2 p_g,\\
		\frac{\text{d}}{\text{d}t}p_{e} 
		= &\kappa_{\rm nl} \left(1-6n_{\text{th},w}\right) \left|\beta' \right|^2 p_g
	\end{align}
\end{subequations}
with
\begin{equation*}
	\beta'=-\frac{2\varepsilon_b'}{\kappa_{\rm nl} \left(1-4n_{\text{th},w}\right) + \kappa_b}.
\end{equation*}
Finally, we obtain the qubit evolution
\begin{equation}
	p_e(t) = \frac{1-6n_{\text{th},w}}{1-4n_{\text{th},w}}  \left(  1- \exp\left(-\eta_c' \left| b_{\rm in} \right|^2 t\right) \right)
	\label{sc_mw_circuit:eq:SPDQubitThermal}
\end{equation}
with the modified conversion efficiency
\begin{equation}
	\eta_c' : = \frac{4\kappa_b\kappa_{\rm nl} \left(1-4n_{\text{th},w}\right)}{(\kappa_{\rm nl} \left(1-4n_{\text{th},w}\right)+\kappa_{b})^2}.
\end{equation}
We clearly see that the qubit population is limited by the waste photon number $n_{\text{th},w}$ and can never reach unity.

\bibliography{bibdic.bib}  

\begin{thebibliography}{65}%
\makeatletter
\providecommand \@ifxundefined [1]{%
 \@ifx{#1\undefined}
}%
\providecommand \@ifnum [1]{%
 \ifnum #1\expandafter \@firstoftwo
 \else \expandafter \@secondoftwo
 \fi
}%
\providecommand \@ifx [1]{%
 \ifx #1\expandafter \@firstoftwo
 \else \expandafter \@secondoftwo
 \fi
}%
\providecommand \natexlab [1]{#1}%
\providecommand \enquote  [1]{``#1''}%
\providecommand \bibnamefont  [1]{#1}%
\providecommand \bibfnamefont [1]{#1}%
\providecommand \citenamefont [1]{#1}%
\providecommand \href@noop [0]{\@secondoftwo}%
\providecommand \href [0]{\begingroup \@sanitize@url \@href}%
\providecommand \@href[1]{\@@startlink{#1}\@@href}%
\providecommand \@@href[1]{\endgroup#1\@@endlink}%
\providecommand \@sanitize@url [0]{\catcode `\\12\catcode `\$12\catcode
  `\&12\catcode `\#12\catcode `\^12\catcode `\_12\catcode `\%12\relax}%
\providecommand \@@startlink[1]{}%
\providecommand \@@endlink[0]{}%
\providecommand \url  [0]{\begingroup\@sanitize@url \@url }%
\providecommand \@url [1]{\endgroup\@href {#1}{\urlprefix }}%
\providecommand \urlprefix  [0]{URL }%
\providecommand \Eprint [0]{\href }%
\providecommand \doibase [0]{https://doi.org/}%
\providecommand \selectlanguage [0]{\@gobble}%
\providecommand \bibinfo  [0]{\@secondoftwo}%
\providecommand \bibfield  [0]{\@secondoftwo}%
\providecommand \translation [1]{[#1]}%
\providecommand \BibitemOpen [0]{}%
\providecommand \bibitemStop [0]{}%
\providecommand \bibitemNoStop [0]{.\EOS\space}%
\providecommand \EOS [0]{\spacefactor3000\relax}%
\providecommand \BibitemShut  [1]{\csname bibitem#1\endcsname}%
\let\auto@bib@innerbib\@empty
\bibitem [{\citenamefont {Yurke}\ \emph {et~al.}(1989)\citenamefont {Yurke},
  \citenamefont {Corruccini}, \citenamefont {Kaminsky}, \citenamefont {Rupp},
  \citenamefont {Smith}, \citenamefont {Silver}, \citenamefont {Simon},\ and\
  \citenamefont {Whittaker}}]{Yurke1989}%
  \BibitemOpen
  \bibfield  {author} {\bibinfo {author} {\bibfnamefont {B.}~\bibnamefont
  {Yurke}}, \bibinfo {author} {\bibfnamefont {L.~R.}\ \bibnamefont
  {Corruccini}}, \bibinfo {author} {\bibfnamefont {P.~G.}\ \bibnamefont
  {Kaminsky}}, \bibinfo {author} {\bibfnamefont {L.~W.}\ \bibnamefont {Rupp}},
  \bibinfo {author} {\bibfnamefont {A.~D.}\ \bibnamefont {Smith}}, \bibinfo
  {author} {\bibfnamefont {A.~H.}\ \bibnamefont {Silver}}, \bibinfo {author}
  {\bibfnamefont {R.~W.}\ \bibnamefont {Simon}},\ and\ \bibinfo {author}
  {\bibfnamefont {E.~A.}\ \bibnamefont {Whittaker}},\ }\bibfield  {title}
  {\bibinfo {title} {\textit{Observation of parametric amplification and
  deamplification in a Josephson parametric amplifier}},\ }\href
  {https://doi.org/10.1103/PhysRevA.39.2519} {\bibfield  {journal} {\bibinfo
  {journal} {Phys. Rev. A}\ }\textbf {\bibinfo {volume} {\textbf{39}}},\
  \bibinfo {pages} {2519} (\bibinfo {year} {1989})}\BibitemShut {NoStop}%
\bibitem [{\citenamefont {Castellanos-Beltran}\ \emph
  {et~al.}(2008)\citenamefont {Castellanos-Beltran}, \citenamefont {Irwin},
  \citenamefont {Hilton}, \citenamefont {Vale},\ and\ \citenamefont
  {Lehnert}}]{Castellanos-Beltran2008}%
  \BibitemOpen
  \bibfield  {author} {\bibinfo {author} {\bibfnamefont {M.~A.}\ \bibnamefont
  {Castellanos-Beltran}}, \bibinfo {author} {\bibfnamefont {K.~D.}\
  \bibnamefont {Irwin}}, \bibinfo {author} {\bibfnamefont {G.~C.}\ \bibnamefont
  {Hilton}}, \bibinfo {author} {\bibfnamefont {L.~R.}\ \bibnamefont {Vale}},\
  and\ \bibinfo {author} {\bibfnamefont {K.~W.}\ \bibnamefont {Lehnert}},\
  }\bibfield  {title} {\bibinfo {title} {\textit{Amplification and squeezing of
  quantum noise with a tunable Josephson metamaterial}},\ }\href
  {https://doi.org/10.1038/nphys1090} {\bibfield  {journal} {\bibinfo
  {journal} {Nat. Phys.}\ }\textbf {\bibinfo {volume} {\textbf{4}}},\ \bibinfo
  {pages} {928} (\bibinfo {year} {2008})}\BibitemShut {NoStop}%
\bibitem [{\citenamefont {Bergeal}\ \emph {et~al.}(2010)\citenamefont
  {Bergeal}, \citenamefont {Schackert}, \citenamefont {Metcalfe}, \citenamefont
  {Vijay}, \citenamefont {Manucharyan}, \citenamefont {Frunzio}, \citenamefont
  {Prober}, \citenamefont {Schoelkopf}, \citenamefont {Girvin},\ and\
  \citenamefont {Devoret}}]{Bergeal2010}%
  \BibitemOpen
  \bibfield  {author} {\bibinfo {author} {\bibfnamefont {N.}~\bibnamefont
  {Bergeal}}, \bibinfo {author} {\bibfnamefont {F.}~\bibnamefont {Schackert}},
  \bibinfo {author} {\bibfnamefont {M.}~\bibnamefont {Metcalfe}}, \bibinfo
  {author} {\bibfnamefont {R.}~\bibnamefont {Vijay}}, \bibinfo {author}
  {\bibfnamefont {V.~E.}\ \bibnamefont {Manucharyan}}, \bibinfo {author}
  {\bibfnamefont {L.}~\bibnamefont {Frunzio}}, \bibinfo {author} {\bibfnamefont
  {D.~E.}\ \bibnamefont {Prober}}, \bibinfo {author} {\bibfnamefont {R.~J.}\
  \bibnamefont {Schoelkopf}}, \bibinfo {author} {\bibfnamefont {S.~M.}\
  \bibnamefont {Girvin}},\ and\ \bibinfo {author} {\bibfnamefont {M.~H.}\
  \bibnamefont {Devoret}},\ }\bibfield  {title} {\bibinfo {title}
  {\textit{Phase-preserving amplification near the quantum limit with a
  Josephson ring modulator}},\ }\href {https://doi.org/10.1038/nature09035}
  {\bibfield  {journal} {\bibinfo  {journal} {\textup{Nature}}\ }\textbf
  {\bibinfo {volume} {\textbf{465}}},\ \bibinfo {pages} {64} (\bibinfo {year}
  {2010})}\BibitemShut {NoStop}%
\bibitem [{\citenamefont {Abdo}\ \emph {et~al.}(2014)\citenamefont {Abdo},
  \citenamefont {Sliwa}, \citenamefont {Frunzio},\ and\ \citenamefont
  {Devoret}}]{Abdo2014}%
  \BibitemOpen
  \bibfield  {author} {\bibinfo {author} {\bibfnamefont {B.}~\bibnamefont
  {Abdo}}, \bibinfo {author} {\bibfnamefont {K.}~\bibnamefont {Sliwa}},
  \bibinfo {author} {\bibfnamefont {L.}~\bibnamefont {Frunzio}},\ and\ \bibinfo
  {author} {\bibfnamefont {M.}~\bibnamefont {Devoret}},\ }\bibfield  {title}
  {\bibinfo {title} {\textit{Directional amplification with a Josephson
  circuit}},\ }\href {https://doi.org/10.1103/PhysRevX.3.031001} {\bibfield
  {journal} {\bibinfo  {journal} {\textup{Phys. Rev. X}}\ }\textbf {\bibinfo
  {volume} {\textbf{3}}},\ \bibinfo {pages} {3} (\bibinfo {year}
  {2014})}\BibitemShut {NoStop}%
\bibitem [{\citenamefont {Malnou}\ \emph {et~al.}(2021)\citenamefont {Malnou},
  \citenamefont {Vissers}, \citenamefont {Wheeler}, \citenamefont {Aumentado},
  \citenamefont {Hubmayr}, \citenamefont {Ullom},\ and\ \citenamefont
  {Gao}}]{Malnou2021}%
  \BibitemOpen
  \bibfield  {author} {\bibinfo {author} {\bibfnamefont {M.}~\bibnamefont
  {Malnou}}, \bibinfo {author} {\bibfnamefont {M.~R.}\ \bibnamefont {Vissers}},
  \bibinfo {author} {\bibfnamefont {J.~D.}\ \bibnamefont {Wheeler}}, \bibinfo
  {author} {\bibfnamefont {J.}~\bibnamefont {Aumentado}}, \bibinfo {author}
  {\bibfnamefont {J.}~\bibnamefont {Hubmayr}}, \bibinfo {author} {\bibfnamefont
  {J.~N.}\ \bibnamefont {Ullom}},\ and\ \bibinfo {author} {\bibfnamefont
  {J.}~\bibnamefont {Gao}},\ }\bibfield  {title} {\bibinfo {title}
  {\textit{Three-Wave Mixing Kinetic Inductance Traveling-Wave Amplifier with
  Near-Quantum-Limited Noise Performance}},\ }\href
  {https://doi.org/10.1103/PRXQuantum.2.010302} {\bibfield  {journal} {\bibinfo
   {journal} {PRX Quantum}\ }\textbf {\bibinfo {volume} {\textbf{2}}},\
  \bibinfo {pages} {1} (\bibinfo {year} {2021})}\BibitemShut {NoStop}%
\bibitem [{\citenamefont {Qiu}\ \emph {et~al.}(2023)\citenamefont {Qiu},
  \citenamefont {Grimsmo}, \citenamefont {Peng}, \citenamefont {Kannan},
  \citenamefont {Lienhard}, \citenamefont {Sung}, \citenamefont {Krantz},
  \citenamefont {Bolkhovsky}, \citenamefont {Calusine}, \citenamefont {Kim},
  \citenamefont {Melville}, \citenamefont {Niedzielski}, \citenamefont {Yoder},
  \citenamefont {Schwartz}, \citenamefont {Orlando}, \citenamefont {Siddiqi},
  \citenamefont {Gustavsson}, \citenamefont {Brien},\ and\ \citenamefont
  {Oliver}}]{Qiu2023}%
  \BibitemOpen
  \bibfield  {author} {\bibinfo {author} {\bibfnamefont {J.~Y.}\ \bibnamefont
  {Qiu}}, \bibinfo {author} {\bibfnamefont {A.}~\bibnamefont {Grimsmo}},
  \bibinfo {author} {\bibfnamefont {K.}~\bibnamefont {Peng}}, \bibinfo {author}
  {\bibfnamefont {B.}~\bibnamefont {Kannan}}, \bibinfo {author} {\bibfnamefont
  {B.}~\bibnamefont {Lienhard}}, \bibinfo {author} {\bibfnamefont
  {Y.}~\bibnamefont {Sung}}, \bibinfo {author} {\bibfnamefont {P.}~\bibnamefont
  {Krantz}}, \bibinfo {author} {\bibfnamefont {V.}~\bibnamefont {Bolkhovsky}},
  \bibinfo {author} {\bibfnamefont {G.}~\bibnamefont {Calusine}}, \bibinfo
  {author} {\bibfnamefont {D.}~\bibnamefont {Kim}}, \bibinfo {author}
  {\bibfnamefont {A.}~\bibnamefont {Melville}}, \bibinfo {author}
  {\bibfnamefont {B.~M.}\ \bibnamefont {Niedzielski}}, \bibinfo {author}
  {\bibfnamefont {J.}~\bibnamefont {Yoder}}, \bibinfo {author} {\bibfnamefont
  {M.~E.}\ \bibnamefont {Schwartz}}, \bibinfo {author} {\bibfnamefont {T.~P.}\
  \bibnamefont {Orlando}}, \bibinfo {author} {\bibfnamefont {I.}~\bibnamefont
  {Siddiqi}}, \bibinfo {author} {\bibfnamefont {S.}~\bibnamefont {Gustavsson}},
  \bibinfo {author} {\bibfnamefont {K.~P.~O.}\ \bibnamefont {Brien}},\ and\
  \bibinfo {author} {\bibfnamefont {W.~D.}\ \bibnamefont {Oliver}},\ }\bibfield
   {title} {\bibinfo {title} {\textit{Broadband squeezed microwaves and
  amplification with a Josephson travelling-wave parametric amplifier}},\
  }\href {https://doi.org/10.1038/s41567-022-01929-w} {\bibfield  {journal}
  {\bibinfo  {journal} {Nat. Phys.}\ }\textbf {\bibinfo {volume}
  {\textbf{19}}},\ \bibinfo {pages} {706} (\bibinfo {year} {2023})}\BibitemShut
  {NoStop}%
\bibitem [{\citenamefont {Abdo}\ \emph {et~al.}(2011)\citenamefont {Abdo},
  \citenamefont {Schackert}, \citenamefont {Hatridge}, \citenamefont
  {Rigetti},\ and\ \citenamefont {Devoret}}]{Abdo2011}%
  \BibitemOpen
  \bibfield  {author} {\bibinfo {author} {\bibfnamefont {B.}~\bibnamefont
  {Abdo}}, \bibinfo {author} {\bibfnamefont {F.}~\bibnamefont {Schackert}},
  \bibinfo {author} {\bibfnamefont {M.}~\bibnamefont {Hatridge}}, \bibinfo
  {author} {\bibfnamefont {C.}~\bibnamefont {Rigetti}},\ and\ \bibinfo {author}
  {\bibfnamefont {M.}~\bibnamefont {Devoret}},\ }\bibfield  {title} {\bibinfo
  {title} {\textit{{Josephson amplifier for qubit readout}}},\ }\href
  {https://doi.org/10.1063/1.3653473} {\bibfield  {journal} {\bibinfo
  {journal} {\textup{Appl. Phys. Lett.}}\ }\textbf {\bibinfo {volume}
  {\textbf{99}}},\ \bibinfo {pages} {16} (\bibinfo {year} {2011})}\BibitemShut
  {NoStop}%
\bibitem [{\citenamefont {Kamal}\ \emph {et~al.}(2014)\citenamefont {Kamal},
  \citenamefont {Roy}, \citenamefont {Clarke},\ and\ \citenamefont
  {Devoret}}]{Kamal2014}%
  \BibitemOpen
  \bibfield  {author} {\bibinfo {author} {\bibfnamefont {A.}~\bibnamefont
  {Kamal}}, \bibinfo {author} {\bibfnamefont {A.}~\bibnamefont {Roy}}, \bibinfo
  {author} {\bibfnamefont {J.}~\bibnamefont {Clarke}},\ and\ \bibinfo {author}
  {\bibfnamefont {M.~H.}\ \bibnamefont {Devoret}},\ }\bibfield  {title}
  {\bibinfo {title} {\textit{Asymmetric frequency conversion in nonlinear
  systems driven by a biharmonic pump}},\ }\href
  {https://doi.org/10.1103/PhysRevLett.113.247003} {\bibfield  {journal}
  {\bibinfo  {journal} {\textup{Phys. Rev. Lett.}}\ }\textbf {\bibinfo {volume}
  {\textbf{113}}},\ \bibinfo {pages} {247003} (\bibinfo {year}
  {2014})}\BibitemShut {NoStop}%
\bibitem [{\citenamefont {Jiang}\ \emph {et~al.}(2023)\citenamefont {Jiang},
  \citenamefont {Ruddy}, \citenamefont {Quinlan}, \citenamefont {Malnou},
  \citenamefont {Frattini},\ and\ \citenamefont {Lehnert}}]{Jiang2023}%
  \BibitemOpen
  \bibfield  {author} {\bibinfo {author} {\bibfnamefont {Y.}~\bibnamefont
  {Jiang}}, \bibinfo {author} {\bibfnamefont {E.~P.}\ \bibnamefont {Ruddy}},
  \bibinfo {author} {\bibfnamefont {K.~O.}\ \bibnamefont {Quinlan}}, \bibinfo
  {author} {\bibfnamefont {M.}~\bibnamefont {Malnou}}, \bibinfo {author}
  {\bibfnamefont {N.~E.}\ \bibnamefont {Frattini}},\ and\ \bibinfo {author}
  {\bibfnamefont {K.~W.}\ \bibnamefont {Lehnert}},\ }\bibfield  {title}
  {\bibinfo {title} {\textit{Accelerated Weak Signal Search Using Mode
  Entanglement and State Swapping}},\ }\href
  {https://doi.org/10.1103/PRXQuantum.4.020302} {\bibfield  {journal} {\bibinfo
   {journal} {Phys. Rev. Appl.}\ }\textbf {\bibinfo {volume} {\textbf{10}}},\
  \bibinfo {pages} {1} (\bibinfo {year} {2023})}\BibitemShut {NoStop}%
\bibitem [{\citenamefont {Pechal}\ \emph {et~al.}(2014)\citenamefont {Pechal},
  \citenamefont {Huthmacher}, \citenamefont {Eichler}, \citenamefont {Zeytino},
  \citenamefont {Abdumalikov}, \citenamefont {Berger}, \citenamefont
  {Wallraff},\ and\ \citenamefont {Filipp}}]{Pechal2014}%
  \BibitemOpen
  \bibfield  {author} {\bibinfo {author} {\bibfnamefont {M.}~\bibnamefont
  {Pechal}}, \bibinfo {author} {\bibfnamefont {L.}~\bibnamefont {Huthmacher}},
  \bibinfo {author} {\bibfnamefont {C.}~\bibnamefont {Eichler}}, \bibinfo
  {author} {\bibfnamefont {S.}~\bibnamefont {Zeytino}}, \bibinfo {author}
  {\bibfnamefont {A.~A.}\ \bibnamefont {Abdumalikov}}, \bibinfo {author}
  {\bibfnamefont {S.}~\bibnamefont {Berger}}, \bibinfo {author} {\bibfnamefont
  {A.}~\bibnamefont {Wallraff}},\ and\ \bibinfo {author} {\bibfnamefont
  {S.}~\bibnamefont {Filipp}},\ }\bibfield  {title} {\bibinfo {title}
  {\textit{Microwave-Controlled Generation of Shaped Single Photons in Circuit
  Quantum Electrodynamics}},\ }\href
  {https://doi.org/10.1103/PhysRevX.4.041010} {\bibfield  {journal} {\bibinfo
  {journal} {Phys. Rev. X}\ }\textbf {\bibinfo {volume} {\textbf{4}}},\
  \bibinfo {pages} {041010} (\bibinfo {year} {2014})}\BibitemShut {NoStop}%
\bibitem [{\citenamefont {Flurin}\ \emph {et~al.}(2012)\citenamefont {Flurin},
  \citenamefont {Roch}, \citenamefont {Mallet}, \citenamefont {Devoret},\ and\
  \citenamefont {Huard}}]{Flurin2012}%
  \BibitemOpen
  \bibfield  {author} {\bibinfo {author} {\bibfnamefont {E.}~\bibnamefont
  {Flurin}}, \bibinfo {author} {\bibfnamefont {N.}~\bibnamefont {Roch}},
  \bibinfo {author} {\bibfnamefont {F.}~\bibnamefont {Mallet}}, \bibinfo
  {author} {\bibfnamefont {M.~H.}\ \bibnamefont {Devoret}},\ and\ \bibinfo
  {author} {\bibfnamefont {B.}~\bibnamefont {Huard}},\ }\bibfield  {title}
  {\bibinfo {title} {\textit{Generating entangled microwave radiation over two
  transmission lines}},\ }\href
  {https://doi.org/10.1103/PhysRevLett.109.183901} {\bibfield  {journal}
  {\bibinfo  {journal} {Phys. Rev. Lett.}\ }\textbf {\bibinfo {volume}
  {\textbf{109}}},\ \bibinfo {pages} {183901} (\bibinfo {year}
  {2012})}\BibitemShut {NoStop}%
\bibitem [{\citenamefont {Murch}\ \emph {et~al.}(2012)\citenamefont {Murch},
  \citenamefont {Vool}, \citenamefont {Zhou}, \citenamefont {Weber},
  \citenamefont {Girvin},\ and\ \citenamefont {Siddiqi}}]{Murch2012}%
  \BibitemOpen
  \bibfield  {author} {\bibinfo {author} {\bibfnamefont {K.~W.}\ \bibnamefont
  {Murch}}, \bibinfo {author} {\bibfnamefont {U.}~\bibnamefont {Vool}},
  \bibinfo {author} {\bibfnamefont {D.}~\bibnamefont {Zhou}}, \bibinfo {author}
  {\bibfnamefont {S.~J.}\ \bibnamefont {Weber}}, \bibinfo {author}
  {\bibfnamefont {S.~M.}\ \bibnamefont {Girvin}},\ and\ \bibinfo {author}
  {\bibfnamefont {I.}~\bibnamefont {Siddiqi}},\ }\bibfield  {title} {\bibinfo
  {title} {\textit{Cavity-assisted quantum bath engineering}},\ }\href
  {https://doi.org/10.1103/PhysRevLett.109.183602} {\bibfield  {journal}
  {\bibinfo  {journal} {Phys. Rev. Lett.}\ }\textbf {\bibinfo {volume}
  {\textbf{109}}},\ \bibinfo {pages} {183602} (\bibinfo {year}
  {2012})}\BibitemShut {NoStop}%
\bibitem [{\citenamefont {Leghtas}\ \emph {et~al.}(2015)\citenamefont
  {Leghtas}, \citenamefont {Touzard}, \citenamefont {Pop}, \citenamefont {Kou},
  \citenamefont {Vlastakis}, \citenamefont {Petrenko}, \citenamefont {Sliwa},
  \citenamefont {Narla}, \citenamefont {Shankar}, \citenamefont {Hatridge},
  \citenamefont {Reagor}, \citenamefont {Frunzio}, \citenamefont {Schoelkopf},
  \citenamefont {Mirrahimi},\ and\ \citenamefont {Devoret}}]{Leghtas2015}%
  \BibitemOpen
  \bibfield  {author} {\bibinfo {author} {\bibfnamefont {Z.}~\bibnamefont
  {Leghtas}}, \bibinfo {author} {\bibfnamefont {S.}~\bibnamefont {Touzard}},
  \bibinfo {author} {\bibfnamefont {I.~M.}\ \bibnamefont {Pop}}, \bibinfo
  {author} {\bibfnamefont {A.}~\bibnamefont {Kou}}, \bibinfo {author}
  {\bibfnamefont {B.}~\bibnamefont {Vlastakis}}, \bibinfo {author}
  {\bibfnamefont {A.}~\bibnamefont {Petrenko}}, \bibinfo {author}
  {\bibfnamefont {K.~M.}\ \bibnamefont {Sliwa}}, \bibinfo {author}
  {\bibfnamefont {A.}~\bibnamefont {Narla}}, \bibinfo {author} {\bibfnamefont
  {S.}~\bibnamefont {Shankar}}, \bibinfo {author} {\bibfnamefont {M.~J.}\
  \bibnamefont {Hatridge}}, \bibinfo {author} {\bibfnamefont {M.}~\bibnamefont
  {Reagor}}, \bibinfo {author} {\bibfnamefont {L.}~\bibnamefont {Frunzio}},
  \bibinfo {author} {\bibfnamefont {R.~J.}\ \bibnamefont {Schoelkopf}},
  \bibinfo {author} {\bibfnamefont {M.}~\bibnamefont {Mirrahimi}},\ and\
  \bibinfo {author} {\bibfnamefont {M.~H.}\ \bibnamefont {Devoret}},\
  }\bibfield  {title} {\bibinfo {title} {\textit{Confining the state of light
  to a quantum manifold by engineered two- photon}},\ }\href
  {https://doi.org/10.1126/science.aaa2085} {\bibfield  {journal} {\bibinfo
  {journal} {Science}\ }\textbf {\bibinfo {volume} {\textbf{347}}},\ \bibinfo
  {pages} {853} (\bibinfo {year} {2015})}\BibitemShut {NoStop}%
\bibitem [{\citenamefont {Dassonneville}\ \emph {et~al.}(2020)\citenamefont
  {Dassonneville}, \citenamefont {Assouly}, \citenamefont {Peronnin},
  \citenamefont {Rouchon},\ and\ \citenamefont {Huard}}]{Dassonneville2020}%
  \BibitemOpen
  \bibfield  {author} {\bibinfo {author} {\bibfnamefont {R.}~\bibnamefont
  {Dassonneville}}, \bibinfo {author} {\bibfnamefont {R.}~\bibnamefont
  {Assouly}}, \bibinfo {author} {\bibfnamefont {T.}~\bibnamefont {Peronnin}},
  \bibinfo {author} {\bibfnamefont {P.}~\bibnamefont {Rouchon}},\ and\ \bibinfo
  {author} {\bibfnamefont {B.}~\bibnamefont {Huard}},\ }\bibfield  {title}
  {\bibinfo {title} {\textit{Number-Resolved Photocounter for Propagating
  Microwave Mode}},\ }\href {https://doi.org/10.1103/PhysRevApplied.14.044022}
  {\bibfield  {journal} {\bibinfo  {journal} {\textup{Phys. Rev. Appl.}}\
  }\textbf {\bibinfo {volume} {\textbf{14}}},\ \bibinfo {pages} {044022}
  (\bibinfo {year} {2020})}\BibitemShut {NoStop}%
\bibitem [{\citenamefont {Lescanne}\ \emph {et~al.}(2020)\citenamefont
  {Lescanne}, \citenamefont {Del{\'{e}}glise}, \citenamefont {Albertinale},
  \citenamefont {R{\'{e}}glade}, \citenamefont {Capelle}, \citenamefont
  {Ivanov}, \citenamefont {Jacqmin}, \citenamefont {Leghtas},\ and\
  \citenamefont {Flurin}}]{Lescanne2020}%
  \BibitemOpen
  \bibfield  {author} {\bibinfo {author} {\bibfnamefont {R.}~\bibnamefont
  {Lescanne}}, \bibinfo {author} {\bibfnamefont {S.}~\bibnamefont
  {Del{\'{e}}glise}}, \bibinfo {author} {\bibfnamefont {E.}~\bibnamefont
  {Albertinale}}, \bibinfo {author} {\bibfnamefont {U.}~\bibnamefont
  {R{\'{e}}glade}}, \bibinfo {author} {\bibfnamefont {T.}~\bibnamefont
  {Capelle}}, \bibinfo {author} {\bibfnamefont {E.}~\bibnamefont {Ivanov}},
  \bibinfo {author} {\bibfnamefont {T.}~\bibnamefont {Jacqmin}}, \bibinfo
  {author} {\bibfnamefont {Z.}~\bibnamefont {Leghtas}},\ and\ \bibinfo {author}
  {\bibfnamefont {E.}~\bibnamefont {Flurin}},\ }\bibfield  {title} {\bibinfo
  {title} {\textit{Irreversible Qubit-Photon Coupling for the Detection of
  Itinerant Microwave Photons}},\ }\href
  {https://doi.org/10.1103/PhysRevX.10.021038} {\bibfield  {journal} {\bibinfo
  {journal} {\textup{Phys. Rev. X}}\ }\textbf {\bibinfo {volume}
  {\textbf{10}}},\ \bibinfo {pages} {021038} (\bibinfo {year}
  {2020})}\BibitemShut {NoStop}%
\bibitem [{\citenamefont {Ma}\ \emph {et~al.}(2012{\natexlab{a}})\citenamefont
  {Ma}, \citenamefont {Herbst}, \citenamefont {Scheidl}, \citenamefont {Wang},
  \citenamefont {Kropatschek}, \citenamefont {Naylor}, \citenamefont
  {Wittmann}, \citenamefont {Mech}, \citenamefont {Kofler}, \citenamefont
  {Anisimova}, \citenamefont {Makarov}, \citenamefont {Jennewein},
  \citenamefont {Ursin},\ and\ \citenamefont {Zeilinger}}]{Ma2012b}%
  \BibitemOpen
  \bibfield  {author} {\bibinfo {author} {\bibfnamefont {X.~S.}\ \bibnamefont
  {Ma}}, \bibinfo {author} {\bibfnamefont {T.}~\bibnamefont {Herbst}}, \bibinfo
  {author} {\bibfnamefont {T.}~\bibnamefont {Scheidl}}, \bibinfo {author}
  {\bibfnamefont {D.}~\bibnamefont {Wang}}, \bibinfo {author} {\bibfnamefont
  {S.}~\bibnamefont {Kropatschek}}, \bibinfo {author} {\bibfnamefont
  {W.}~\bibnamefont {Naylor}}, \bibinfo {author} {\bibfnamefont
  {B.}~\bibnamefont {Wittmann}}, \bibinfo {author} {\bibfnamefont
  {A.}~\bibnamefont {Mech}}, \bibinfo {author} {\bibfnamefont {J.}~\bibnamefont
  {Kofler}}, \bibinfo {author} {\bibfnamefont {E.}~\bibnamefont {Anisimova}},
  \bibinfo {author} {\bibfnamefont {V.}~\bibnamefont {Makarov}}, \bibinfo
  {author} {\bibfnamefont {T.}~\bibnamefont {Jennewein}}, \bibinfo {author}
  {\bibfnamefont {R.}~\bibnamefont {Ursin}},\ and\ \bibinfo {author}
  {\bibfnamefont {A.}~\bibnamefont {Zeilinger}},\ }\bibfield  {title} {\bibinfo
  {title} {\textit{Quantum teleportation over 143 kilometres using active
  feed-forward}},\ }\href {https://doi.org/10.1038/nature11472} {\bibfield
  {journal} {\bibinfo  {journal} {Nature}\ }\textbf {\bibinfo {volume}
  {\textbf{489}}},\ \bibinfo {pages} {269} (\bibinfo {year}
  {2012}{\natexlab{a}})}\BibitemShut {NoStop}%
\bibitem [{\citenamefont {Abdelkhalek}\ \emph {et~al.}(2016)\citenamefont
  {Abdelkhalek}, \citenamefont {Syllwasschy}, \citenamefont {Cerf},
  \citenamefont {Fiur{\'{a}}{\v{s}}ek},\ and\ \citenamefont
  {Schnabel}}]{Abdelkhalek2016}%
  \BibitemOpen
  \bibfield  {author} {\bibinfo {author} {\bibfnamefont {D.}~\bibnamefont
  {Abdelkhalek}}, \bibinfo {author} {\bibfnamefont {M.}~\bibnamefont
  {Syllwasschy}}, \bibinfo {author} {\bibfnamefont {N.~J.}\ \bibnamefont
  {Cerf}}, \bibinfo {author} {\bibfnamefont {J.}~\bibnamefont
  {Fiur{\'{a}}{\v{s}}ek}},\ and\ \bibinfo {author} {\bibfnamefont
  {R.}~\bibnamefont {Schnabel}},\ }\bibfield  {title} {\bibinfo {title}
  {\textit{Efficient entanglement distillation without quantum memory}},\
  }\href {https://doi.org/10.1038/ncomms11720} {\bibfield  {journal} {\bibinfo
  {journal} {Nat. Commun.}\ }\textbf {\bibinfo {volume} {\textbf{7}}},\
  \bibinfo {pages} {11720} (\bibinfo {year} {2016})}\BibitemShut {NoStop}%
\bibitem [{\citenamefont {Ma}\ \emph {et~al.}(2012{\natexlab{b}})\citenamefont
  {Ma}, \citenamefont {Zotter}, \citenamefont {Kofler}, \citenamefont {Ursin},
  \citenamefont {Jennewein}, \citenamefont {Brukner},\ and\ \citenamefont
  {Zeilinger}}]{Ma2012a}%
  \BibitemOpen
  \bibfield  {author} {\bibinfo {author} {\bibfnamefont {X.~S.}\ \bibnamefont
  {Ma}}, \bibinfo {author} {\bibfnamefont {S.}~\bibnamefont {Zotter}}, \bibinfo
  {author} {\bibfnamefont {J.}~\bibnamefont {Kofler}}, \bibinfo {author}
  {\bibfnamefont {R.}~\bibnamefont {Ursin}}, \bibinfo {author} {\bibfnamefont
  {T.}~\bibnamefont {Jennewein}}, \bibinfo {author} {\bibfnamefont
  {{\v{C}}.}~\bibnamefont {Brukner}},\ and\ \bibinfo {author} {\bibfnamefont
  {A.}~\bibnamefont {Zeilinger}},\ }\bibfield  {title} {\bibinfo {title}
  {\textit{Experimental delayed-choice entanglement swapping}},\ }\href
  {https://doi.org/10.1038/nphys2294} {\bibfield  {journal} {\bibinfo
  {journal} {Nat. Phys.}\ }\textbf {\bibinfo {volume} {\textbf{8}}},\ \bibinfo
  {pages} {479} (\bibinfo {year} {2012}{\natexlab{b}})}\BibitemShut {NoStop}%
\bibitem [{\citenamefont {Scully}\ and\ \citenamefont
  {Dr{\"{u}}hl}(1982)}]{Scully1982}%
  \BibitemOpen
  \bibfield  {author} {\bibinfo {author} {\bibfnamefont {M.~O.}\ \bibnamefont
  {Scully}}\ and\ \bibinfo {author} {\bibfnamefont {K.}~\bibnamefont
  {Dr{\"{u}}hl}},\ }\bibfield  {title} {\bibinfo {title} {\textit{Quantum
  eraser: A proposed photon correlation experiment concerning observation and
  "delayed choice" in quantum mechanics}},\ }\href
  {https://doi.org/10.1103/PhysRevA.25.2208} {\bibfield  {journal} {\bibinfo
  {journal} {Phys. Rev. A}\ }\textbf {\bibinfo {volume} {\textbf{25}}},\
  \bibinfo {pages} {2208} (\bibinfo {year} {1982})}\BibitemShut {NoStop}%
\bibitem [{\citenamefont {Yuan}\ \emph {et~al.}(2008)\citenamefont {Yuan},
  \citenamefont {Chen}, \citenamefont {Zhao}, \citenamefont {Chen},
  \citenamefont {Schmiedmayer},\ and\ \citenamefont {Pan}}]{Yuan2008}%
  \BibitemOpen
  \bibfield  {author} {\bibinfo {author} {\bibfnamefont {Z.~S.}\ \bibnamefont
  {Yuan}}, \bibinfo {author} {\bibfnamefont {Y.~A.}\ \bibnamefont {Chen}},
  \bibinfo {author} {\bibfnamefont {B.}~\bibnamefont {Zhao}}, \bibinfo {author}
  {\bibfnamefont {S.}~\bibnamefont {Chen}}, \bibinfo {author} {\bibfnamefont
  {J.}~\bibnamefont {Schmiedmayer}},\ and\ \bibinfo {author} {\bibfnamefont
  {J.~W.}\ \bibnamefont {Pan}},\ }\bibfield  {title} {\bibinfo {title}
  {\textit{Experimental demonstration of a BDCZ quantum repeater node}},\
  }\href {https://doi.org/10.1038/nature07241} {\bibfield  {journal} {\bibinfo
  {journal} {Nature}\ }\textbf {\bibinfo {volume} {\textbf{454}}},\ \bibinfo
  {pages} {1098} (\bibinfo {year} {2008})}\BibitemShut {NoStop}%
\bibitem [{\citenamefont {Deng}\ \emph {et~al.}(2023)\citenamefont {Deng},
  \citenamefont {Gong}, \citenamefont {Gu}, \citenamefont {Zhang},
  \citenamefont {Liu}, \citenamefont {Su}, \citenamefont {Tang}, \citenamefont
  {Xu}, \citenamefont {Jia}, \citenamefont {Chen}, \citenamefont {Zhong},
  \citenamefont {Wang}, \citenamefont {Yan}, \citenamefont {Hu}, \citenamefont
  {Huang}, \citenamefont {Zhang}, \citenamefont {Li}, \citenamefont {Jiang},
  \citenamefont {You}, \citenamefont {Wang}, \citenamefont {Li}, \citenamefont
  {Liu}, \citenamefont {Lu},\ and\ \citenamefont {Pan}}]{Deng2023}%
  \BibitemOpen
  \bibfield  {author} {\bibinfo {author} {\bibfnamefont {Y.~H.}\ \bibnamefont
  {Deng}}, \bibinfo {author} {\bibfnamefont {S.~Q.}\ \bibnamefont {Gong}},
  \bibinfo {author} {\bibfnamefont {Y.~C.}\ \bibnamefont {Gu}}, \bibinfo
  {author} {\bibfnamefont {Z.~J.}\ \bibnamefont {Zhang}}, \bibinfo {author}
  {\bibfnamefont {H.~L.}\ \bibnamefont {Liu}}, \bibinfo {author} {\bibfnamefont
  {H.}~\bibnamefont {Su}}, \bibinfo {author} {\bibfnamefont {H.~Y.}\
  \bibnamefont {Tang}}, \bibinfo {author} {\bibfnamefont {J.~M.}\ \bibnamefont
  {Xu}}, \bibinfo {author} {\bibfnamefont {M.~H.}\ \bibnamefont {Jia}},
  \bibinfo {author} {\bibfnamefont {M.~C.}\ \bibnamefont {Chen}}, \bibinfo
  {author} {\bibfnamefont {H.~S.}\ \bibnamefont {Zhong}}, \bibinfo {author}
  {\bibfnamefont {H.}~\bibnamefont {Wang}}, \bibinfo {author} {\bibfnamefont
  {J.}~\bibnamefont {Yan}}, \bibinfo {author} {\bibfnamefont {Y.}~\bibnamefont
  {Hu}}, \bibinfo {author} {\bibfnamefont {J.}~\bibnamefont {Huang}}, \bibinfo
  {author} {\bibfnamefont {W.~J.}\ \bibnamefont {Zhang}}, \bibinfo {author}
  {\bibfnamefont {H.}~\bibnamefont {Li}}, \bibinfo {author} {\bibfnamefont
  {X.}~\bibnamefont {Jiang}}, \bibinfo {author} {\bibfnamefont
  {L.}~\bibnamefont {You}}, \bibinfo {author} {\bibfnamefont {Z.}~\bibnamefont
  {Wang}}, \bibinfo {author} {\bibfnamefont {L.}~\bibnamefont {Li}}, \bibinfo
  {author} {\bibfnamefont {N.~L.}\ \bibnamefont {Liu}}, \bibinfo {author}
  {\bibfnamefont {C.~Y.}\ \bibnamefont {Lu}},\ and\ \bibinfo {author}
  {\bibfnamefont {J.~W.}\ \bibnamefont {Pan}},\ }\bibfield  {title} {\bibinfo
  {title} {\textit{Solving Graph Problems Using Gaussian Boson Sampling}},\
  }\href {https://doi.org/10.1103/PhysRevLett.130.190601} {\bibfield  {journal}
  {\bibinfo  {journal} {Phys. Rev. Lett.}\ }\textbf {\bibinfo {volume}
  {\textbf{130}}},\ \bibinfo {pages} {190601} (\bibinfo {year}
  {2023})}\BibitemShut {NoStop}%
\bibitem [{\citenamefont {Jin}\ \emph {et~al.}(2012)\citenamefont {Jin},
  \citenamefont {Yi}, \citenamefont {Yang}, \citenamefont {Zhou}, \citenamefont
  {Yang},\ and\ \citenamefont {Peng}}]{Jin2012}%
  \BibitemOpen
  \bibfield  {author} {\bibinfo {author} {\bibfnamefont {X.~M.}\ \bibnamefont
  {Jin}}, \bibinfo {author} {\bibfnamefont {Z.~H.}\ \bibnamefont {Yi}},
  \bibinfo {author} {\bibfnamefont {B.}~\bibnamefont {Yang}}, \bibinfo {author}
  {\bibfnamefont {F.}~\bibnamefont {Zhou}}, \bibinfo {author} {\bibfnamefont
  {T.}~\bibnamefont {Yang}},\ and\ \bibinfo {author} {\bibfnamefont {C.~Z.}\
  \bibnamefont {Peng}},\ }\bibfield  {title} {\bibinfo {title}
  {\textit{Experimental quantum error detection}},\ }\href
  {https://doi.org/10.1038/srep00626} {\bibfield  {journal} {\bibinfo
  {journal} {Sci. Rep.}\ }\textbf {\bibinfo {volume} {\textbf{2}}},\ \bibinfo
  {pages} {626} (\bibinfo {year} {2012})}\BibitemShut {NoStop}%
\bibitem [{\citenamefont {Chen}\ \emph {et~al.}(2011)\citenamefont {Chen},
  \citenamefont {Hover}, \citenamefont {Sendelbach}, \citenamefont {Maurer},
  \citenamefont {Merkel}, \citenamefont {Pritchett}, \citenamefont {Wilhelm},\
  and\ \citenamefont {McDermott}}]{Chen2011}%
  \BibitemOpen
  \bibfield  {author} {\bibinfo {author} {\bibfnamefont {Y.~F.}\ \bibnamefont
  {Chen}}, \bibinfo {author} {\bibfnamefont {D.}~\bibnamefont {Hover}},
  \bibinfo {author} {\bibfnamefont {S.}~\bibnamefont {Sendelbach}}, \bibinfo
  {author} {\bibfnamefont {L.}~\bibnamefont {Maurer}}, \bibinfo {author}
  {\bibfnamefont {S.~T.}\ \bibnamefont {Merkel}}, \bibinfo {author}
  {\bibfnamefont {E.~J.}\ \bibnamefont {Pritchett}}, \bibinfo {author}
  {\bibfnamefont {F.~K.}\ \bibnamefont {Wilhelm}},\ and\ \bibinfo {author}
  {\bibfnamefont {R.}~\bibnamefont {McDermott}},\ }\bibfield  {title} {\bibinfo
  {title} {\textit{Microwave photon counter based on josephson junctions}},\
  }\href {https://doi.org/10.1103/PhysRevLett.107.217401} {\bibfield  {journal}
  {\bibinfo  {journal} {Phys. Rev. Lett.}\ }\textbf {\bibinfo {volume}
  {\textbf{107}}},\ \bibinfo {pages} {217401} (\bibinfo {year}
  {2011})}\BibitemShut {NoStop}%
\bibitem [{\citenamefont {Inomata}\ \emph {et~al.}(2016)\citenamefont
  {Inomata}, \citenamefont {Lin}, \citenamefont {Koshino}, \citenamefont
  {Oliver}, \citenamefont {Tsai}, \citenamefont {Yamamoto},\ and\ \citenamefont
  {Nakamura}}]{Inomata2016}%
  \BibitemOpen
  \bibfield  {author} {\bibinfo {author} {\bibfnamefont {K.}~\bibnamefont
  {Inomata}}, \bibinfo {author} {\bibfnamefont {Z.}~\bibnamefont {Lin}},
  \bibinfo {author} {\bibfnamefont {K.}~\bibnamefont {Koshino}}, \bibinfo
  {author} {\bibfnamefont {W.~D.}\ \bibnamefont {Oliver}}, \bibinfo {author}
  {\bibfnamefont {J.-S.}\ \bibnamefont {Tsai}}, \bibinfo {author}
  {\bibfnamefont {T.}~\bibnamefont {Yamamoto}},\ and\ \bibinfo {author}
  {\bibfnamefont {Y.}~\bibnamefont {Nakamura}},\ }\bibfield  {title} {\bibinfo
  {title} {\textit{Single microwave-photon detector using an artiﬁcial L-type
  three-level system}},\ }\href {https://doi.org/10.1038/ncomms12303}
  {\bibfield  {journal} {\bibinfo  {journal} {Nat. Commun.}\ }\textbf {\bibinfo
  {volume} {\textbf{7}}},\ \bibinfo {pages} {12303} (\bibinfo {year}
  {2016})}\BibitemShut {NoStop}%
\bibitem [{\citenamefont {Kono}\ \emph {et~al.}(2018)\citenamefont {Kono},
  \citenamefont {Koshino}, \citenamefont {Tabuchi}, \citenamefont {Noguchi},\
  and\ \citenamefont {Nakamura}}]{Kono2018}%
  \BibitemOpen
  \bibfield  {author} {\bibinfo {author} {\bibfnamefont {S.}~\bibnamefont
  {Kono}}, \bibinfo {author} {\bibfnamefont {K.}~\bibnamefont {Koshino}},
  \bibinfo {author} {\bibfnamefont {Y.}~\bibnamefont {Tabuchi}}, \bibinfo
  {author} {\bibfnamefont {A.}~\bibnamefont {Noguchi}},\ and\ \bibinfo {author}
  {\bibfnamefont {Y.}~\bibnamefont {Nakamura}},\ }\bibfield  {title} {\bibinfo
  {title} {\textit{Quantum non-demolition detection of an itinerant microwave
  photon}},\ }\href {https://doi.org/10.1038/s41567-018-0066-3} {\bibfield
  {journal} {\bibinfo  {journal} {Nat. Phys.}\ }\textbf {\bibinfo {volume}
  {\textbf{14}}},\ \bibinfo {pages} {546} (\bibinfo {year} {2018})}\BibitemShut
  {NoStop}%
\bibitem [{\citenamefont {Besse}\ \emph {et~al.}(2018)\citenamefont {Besse},
  \citenamefont {Gasparinetti}, \citenamefont {Collodo}, \citenamefont
  {Walter}, \citenamefont {Kurpiers}, \citenamefont {Pechal}, \citenamefont
  {Eichler},\ and\ \citenamefont {Wallraff}}]{Besse2018}%
  \BibitemOpen
  \bibfield  {author} {\bibinfo {author} {\bibfnamefont {J.~C.}\ \bibnamefont
  {Besse}}, \bibinfo {author} {\bibfnamefont {S.}~\bibnamefont {Gasparinetti}},
  \bibinfo {author} {\bibfnamefont {M.~C.}\ \bibnamefont {Collodo}}, \bibinfo
  {author} {\bibfnamefont {T.}~\bibnamefont {Walter}}, \bibinfo {author}
  {\bibfnamefont {P.}~\bibnamefont {Kurpiers}}, \bibinfo {author}
  {\bibfnamefont {M.}~\bibnamefont {Pechal}}, \bibinfo {author} {\bibfnamefont
  {C.}~\bibnamefont {Eichler}},\ and\ \bibinfo {author} {\bibfnamefont
  {A.}~\bibnamefont {Wallraff}},\ }\bibfield  {title} {\bibinfo {title}
  {\textit{Single-Shot Quantum Nondemolition Detection of Individual Itinerant
  Microwave Photons}},\ }\href {https://doi.org/10.1103/PhysRevX.8.021003}
  {\bibfield  {journal} {\bibinfo  {journal} {Phys. Rev. X}\ }\textbf {\bibinfo
  {volume} {8}},\ \bibinfo {pages} {021003} (\bibinfo {year}
  {2018})}\BibitemShut {NoStop}%
\bibitem [{\citenamefont {Varnava}\ \emph {et~al.}(2008)\citenamefont
  {Varnava}, \citenamefont {Browne},\ and\ \citenamefont
  {Rudolph}}]{Varnava2008}%
  \BibitemOpen
  \bibfield  {author} {\bibinfo {author} {\bibfnamefont {M.}~\bibnamefont
  {Varnava}}, \bibinfo {author} {\bibfnamefont {D.~E.}\ \bibnamefont
  {Browne}},\ and\ \bibinfo {author} {\bibfnamefont {T.}~\bibnamefont
  {Rudolph}},\ }\bibfield  {title} {\bibinfo {title} {\textit{How good must
  single photon sources and detectors be for efficient linear optical quantum
  computation?}},\ }\href {https://doi.org/10.1103/PhysRevLett.100.060502}
  {\bibfield  {journal} {\bibinfo  {journal} {Phys. Rev. Lett.}\ }\textbf
  {\bibinfo {volume} {100}},\ \bibinfo {pages} {060502} (\bibinfo {year}
  {2008})}\BibitemShut {NoStop}%
\bibitem [{\citenamefont {Hadfield}(2009)}]{Hadfield2009}%
  \BibitemOpen
  \bibfield  {author} {\bibinfo {author} {\bibfnamefont {R.~H.}\ \bibnamefont
  {Hadfield}},\ }\bibfield  {title} {\bibinfo {title} {\textit{Single-photon
  detectors for optical quantum information applications}},\ }\href
  {https://doi.org/10.1038/nphoton.2009.230} {\bibfield  {journal} {\bibinfo
  {journal} {Nat. Photon.}\ }\textbf {\bibinfo {volume} {\textbf{3}}},\
  \bibinfo {pages} {696} (\bibinfo {year} {2009})}\BibitemShut {NoStop}%
\bibitem [{\citenamefont {Lamoreaux}\ \emph {et~al.}(2013)\citenamefont
  {Lamoreaux}, \citenamefont {{Van Bibber}}, \citenamefont {Lehnert},\ and\
  \citenamefont {Carosi}}]{Lamoreaux2013}%
  \BibitemOpen
  \bibfield  {author} {\bibinfo {author} {\bibfnamefont {S.~K.}\ \bibnamefont
  {Lamoreaux}}, \bibinfo {author} {\bibfnamefont {K.~A.}\ \bibnamefont {{Van
  Bibber}}}, \bibinfo {author} {\bibfnamefont {K.~W.}\ \bibnamefont
  {Lehnert}},\ and\ \bibinfo {author} {\bibfnamefont {G.}~\bibnamefont
  {Carosi}},\ }\bibfield  {title} {\bibinfo {title} {\textit{Analysis of
  single-photon and linear amplifier detectors for microwave cavity dark matter
  axion searches}},\ }\href {https://doi.org/10.1103/PhysRevD.88.035020}
  {\bibfield  {journal} {\bibinfo  {journal} {Phys. Rev. D}\ }\textbf {\bibinfo
  {volume} {\textbf{88}}},\ \bibinfo {pages} {035020} (\bibinfo {year}
  {2013})}\BibitemShut {NoStop}%
\bibitem [{\citenamefont {Chunnilall}\ \emph {et~al.}(2014)\citenamefont
  {Chunnilall}, \citenamefont {Degiovanni}, \citenamefont {K{\"{u}}ck},
  \citenamefont {M{\"{u}}ller},\ and\ \citenamefont
  {Sinclair}}]{Chunnilall2014}%
  \BibitemOpen
  \bibfield  {author} {\bibinfo {author} {\bibfnamefont {C.~J.}\ \bibnamefont
  {Chunnilall}}, \bibinfo {author} {\bibfnamefont {I.~P.}\ \bibnamefont
  {Degiovanni}}, \bibinfo {author} {\bibfnamefont {S.}~\bibnamefont
  {K{\"{u}}ck}}, \bibinfo {author} {\bibfnamefont {I.}~\bibnamefont
  {M{\"{u}}ller}},\ and\ \bibinfo {author} {\bibfnamefont {A.~G.}\ \bibnamefont
  {Sinclair}},\ }\bibfield  {title} {\bibinfo {title} {\textit{Metrology of
  single-photon sources and detectors: a review}},\ }\href
  {https://doi.org/10.1117/1.oe.53.8.081910} {\bibfield  {journal} {\bibinfo
  {journal} {Opt. Eng.}\ }\textbf {\bibinfo {volume} {53}},\ \bibinfo {pages}
  {081910} (\bibinfo {year} {2014})}\BibitemShut {NoStop}%
\bibitem [{\citenamefont {Barzanjeh}\ \emph {et~al.}(2015)\citenamefont
  {Barzanjeh}, \citenamefont {Guha}, \citenamefont {Weedbrook}, \citenamefont
  {Vitali}, \citenamefont {Shapiro},\ and\ \citenamefont
  {Pirandola}}]{Barzanjeh2015}%
  \BibitemOpen
  \bibfield  {author} {\bibinfo {author} {\bibfnamefont {S.}~\bibnamefont
  {Barzanjeh}}, \bibinfo {author} {\bibfnamefont {S.}~\bibnamefont {Guha}},
  \bibinfo {author} {\bibfnamefont {C.}~\bibnamefont {Weedbrook}}, \bibinfo
  {author} {\bibfnamefont {D.}~\bibnamefont {Vitali}}, \bibinfo {author}
  {\bibfnamefont {J.~H.}\ \bibnamefont {Shapiro}},\ and\ \bibinfo {author}
  {\bibfnamefont {S.}~\bibnamefont {Pirandola}},\ }\bibfield  {title} {\bibinfo
  {title} {\textit{Microwave quantum illumination}},\ }\href
  {https://doi.org/10.1103/PhysRevLett.114.080503} {\bibfield  {journal}
  {\bibinfo  {journal} {Phys. Rev. Lett.}\ }\textbf {\bibinfo {volume}
  {\textbf{114}}},\ \bibinfo {pages} {080503} (\bibinfo {year}
  {2015})}\BibitemShut {NoStop}%
\bibitem [{\citenamefont {Assouly}\ \emph {et~al.}(2023)\citenamefont
  {Assouly}, \citenamefont {Dassonneville}, \citenamefont {Peronnin},
  \citenamefont {Bienfait},\ and\ \citenamefont {Huard}}]{Assouly2023}%
  \BibitemOpen
  \bibfield  {author} {\bibinfo {author} {\bibfnamefont {R.}~\bibnamefont
  {Assouly}}, \bibinfo {author} {\bibfnamefont {R.}~\bibnamefont
  {Dassonneville}}, \bibinfo {author} {\bibfnamefont {T.}~\bibnamefont
  {Peronnin}}, \bibinfo {author} {\bibfnamefont {A.}~\bibnamefont {Bienfait}},\
  and\ \bibinfo {author} {\bibfnamefont {B.}~\bibnamefont {Huard}},\ }\bibfield
   {title} {\bibinfo {title} {\textit{Quantum advantage in microwave quantum
  radar}},\ }\href {https://doi.org/10.1038/s41567-023-02113-4} {\bibfield
  {journal} {\bibinfo  {journal} {Nat. Phys.}\ }\textbf {\bibinfo {volume}
  {\textbf{19}}},\ \bibinfo {pages} {1418} (\bibinfo {year}
  {2023})}\BibitemShut {NoStop}%
\bibitem [{\citenamefont {Kronowetter}\ \emph {et~al.}(2023)\citenamefont
  {Kronowetter}, \citenamefont {Fesquet}, \citenamefont {Renger}, \citenamefont
  {Honasoge}, \citenamefont {Nojiri}, \citenamefont {Inomata}, \citenamefont
  {Nakamura}, \citenamefont {Marx}, \citenamefont {Gross},\ and\ \citenamefont
  {Fedorov}}]{Kronowetter2023}%
  \BibitemOpen
  \bibfield  {author} {\bibinfo {author} {\bibfnamefont {F.}~\bibnamefont
  {Kronowetter}}, \bibinfo {author} {\bibfnamefont {F.}~\bibnamefont
  {Fesquet}}, \bibinfo {author} {\bibfnamefont {M.}~\bibnamefont {Renger}},
  \bibinfo {author} {\bibfnamefont {K.}~\bibnamefont {Honasoge}}, \bibinfo
  {author} {\bibfnamefont {Y.}~\bibnamefont {Nojiri}}, \bibinfo {author}
  {\bibfnamefont {K.}~\bibnamefont {Inomata}}, \bibinfo {author} {\bibfnamefont
  {Y.}~\bibnamefont {Nakamura}}, \bibinfo {author} {\bibfnamefont
  {A.}~\bibnamefont {Marx}}, \bibinfo {author} {\bibfnamefont {R.}~\bibnamefont
  {Gross}},\ and\ \bibinfo {author} {\bibfnamefont {K.~G.}\ \bibnamefont
  {Fedorov}},\ }\bibfield  {title} {\bibinfo {title} {\textit{Quantum Microwave
  Parametric Interferometer}},\ }\href
  {https://doi.org/10.1103/PhysRevApplied.20.024049} {\bibfield  {journal}
  {\bibinfo  {journal} {Phys. Rev. Appl.}\ }\textbf {\bibinfo {volume}
  {\textbf{20}}},\ \bibinfo {pages} {024049} (\bibinfo {year}
  {2023})}\BibitemShut {NoStop}%
\bibitem [{\citenamefont {Wang}\ \emph {et~al.}(2023)\citenamefont {Wang},
  \citenamefont {Balembois}, \citenamefont {Ran{\v{c}}i{\'{c}}}, \citenamefont
  {Billaud}, \citenamefont {{Le Dantec}}, \citenamefont {Ferrier},
  \citenamefont {Goldner}, \citenamefont {Bertaina}, \citenamefont
  {Chaneli{\`{e}}re}, \citenamefont {Esteve}, \citenamefont {Vion},
  \citenamefont {Bertet},\ and\ \citenamefont {Flurin}}]{Wang2023}%
  \BibitemOpen
  \bibfield  {author} {\bibinfo {author} {\bibfnamefont {Z.}~\bibnamefont
  {Wang}}, \bibinfo {author} {\bibfnamefont {L.}~\bibnamefont {Balembois}},
  \bibinfo {author} {\bibfnamefont {M.}~\bibnamefont {Ran{\v{c}}i{\'{c}}}},
  \bibinfo {author} {\bibfnamefont {E.}~\bibnamefont {Billaud}}, \bibinfo
  {author} {\bibfnamefont {M.}~\bibnamefont {{Le Dantec}}}, \bibinfo {author}
  {\bibfnamefont {A.}~\bibnamefont {Ferrier}}, \bibinfo {author} {\bibfnamefont
  {P.}~\bibnamefont {Goldner}}, \bibinfo {author} {\bibfnamefont
  {S.}~\bibnamefont {Bertaina}}, \bibinfo {author} {\bibfnamefont
  {T.}~\bibnamefont {Chaneli{\`{e}}re}}, \bibinfo {author} {\bibfnamefont
  {D.}~\bibnamefont {Esteve}}, \bibinfo {author} {\bibfnamefont
  {D.}~\bibnamefont {Vion}}, \bibinfo {author} {\bibfnamefont {P.}~\bibnamefont
  {Bertet}},\ and\ \bibinfo {author} {\bibfnamefont {E.}~\bibnamefont
  {Flurin}},\ }\bibfield  {title} {\bibinfo {title} {\textit{Single-electron
  spin resonance detection by microwave photon counting}},\ }\href
  {https://doi.org/10.1038/s41586-023-06097-2} {\bibfield  {journal} {\bibinfo
  {journal} {Nature}\ }\textbf {\bibinfo {volume} {\textbf{619}}},\ \bibinfo
  {pages} {276} (\bibinfo {year} {2023})}\BibitemShut {NoStop}%
\bibitem [{\citenamefont {Billaud}\ \emph {et~al.}(2023)\citenamefont
  {Billaud}, \citenamefont {Balembois}, \citenamefont {{Le Dantec}},
  \citenamefont {Ran{\v{c}}i{\'{c}}}, \citenamefont {Albertinale},
  \citenamefont {Bertaina}, \citenamefont {Chaneli{\`{e}}re}, \citenamefont
  {Goldner}, \citenamefont {Est{\`{e}}ve}, \citenamefont {Vion}, \citenamefont
  {Bertet},\ and\ \citenamefont {Flurin}}]{Billaud2023}%
  \BibitemOpen
  \bibfield  {author} {\bibinfo {author} {\bibfnamefont {E.}~\bibnamefont
  {Billaud}}, \bibinfo {author} {\bibfnamefont {L.}~\bibnamefont {Balembois}},
  \bibinfo {author} {\bibfnamefont {M.}~\bibnamefont {{Le Dantec}}}, \bibinfo
  {author} {\bibfnamefont {M.}~\bibnamefont {Ran{\v{c}}i{\'{c}}}}, \bibinfo
  {author} {\bibfnamefont {E.}~\bibnamefont {Albertinale}}, \bibinfo {author}
  {\bibfnamefont {S.}~\bibnamefont {Bertaina}}, \bibinfo {author}
  {\bibfnamefont {T.}~\bibnamefont {Chaneli{\`{e}}re}}, \bibinfo {author}
  {\bibfnamefont {P.}~\bibnamefont {Goldner}}, \bibinfo {author} {\bibfnamefont
  {D.}~\bibnamefont {Est{\`{e}}ve}}, \bibinfo {author} {\bibfnamefont
  {D.}~\bibnamefont {Vion}}, \bibinfo {author} {\bibfnamefont {P.}~\bibnamefont
  {Bertet}},\ and\ \bibinfo {author} {\bibfnamefont {E.}~\bibnamefont
  {Flurin}},\ }\bibfield  {title} {\bibinfo {title} {\textit{Microwave
  Fluorescence Detection of Spin Echoes}},\ }\href
  {https://doi.org/10.1103/physrevlett.131.100804} {\bibfield  {journal}
  {\bibinfo  {journal} {Phys. Rev. Lett.}\ }\textbf {\bibinfo {volume}
  {\textbf{131}}},\ \bibinfo {pages} {100804} (\bibinfo {year}
  {2023})}\BibitemShut {NoStop}%
\bibitem [{\citenamefont {{Las Heras}}\ \emph {et~al.}(2017)\citenamefont {{Las
  Heras}}, \citenamefont {{Di Candia}}, \citenamefont {Fedorov}, \citenamefont
  {Deppe}, \citenamefont {Sanz},\ and\ \citenamefont {Solano}}]{LasHeras2017}%
  \BibitemOpen
  \bibfield  {author} {\bibinfo {author} {\bibfnamefont {U.}~\bibnamefont {{Las
  Heras}}}, \bibinfo {author} {\bibfnamefont {R.}~\bibnamefont {{Di Candia}}},
  \bibinfo {author} {\bibfnamefont {K.~G.}\ \bibnamefont {Fedorov}}, \bibinfo
  {author} {\bibfnamefont {F.}~\bibnamefont {Deppe}}, \bibinfo {author}
  {\bibfnamefont {M.}~\bibnamefont {Sanz}},\ and\ \bibinfo {author}
  {\bibfnamefont {E.}~\bibnamefont {Solano}},\ }\bibfield  {title} {\bibinfo
  {title} {\textit{Quantum illumination reveals phase-shift inducing
  cloaking}},\ }\href {https://doi.org/10.1038/s41598-017-08505-w} {\bibfield
  {journal} {\bibinfo  {journal} {Sci. Rep.}\ }\textbf {\bibinfo {volume}
  {\textbf{7}}},\ \bibinfo {pages} {9333} (\bibinfo {year} {2017})}\BibitemShut
  {NoStop}%
\bibitem [{\citenamefont {Fedorov}\ \emph {et~al.}(2021)\citenamefont
  {Fedorov}, \citenamefont {Renger}, \citenamefont {Pogorzalek}, \citenamefont
  {{Di Candia}}, \citenamefont {Chen}, \citenamefont {Nojiri}, \citenamefont
  {Inomata}, \citenamefont {Nakamura}, \citenamefont {Partanen}, \citenamefont
  {Marx}, \citenamefont {Gross},\ and\ \citenamefont {Deppe}}]{Fedorov2021}%
  \BibitemOpen
  \bibfield  {author} {\bibinfo {author} {\bibfnamefont {K.~G.}\ \bibnamefont
  {Fedorov}}, \bibinfo {author} {\bibfnamefont {M.}~\bibnamefont {Renger}},
  \bibinfo {author} {\bibfnamefont {S.}~\bibnamefont {Pogorzalek}}, \bibinfo
  {author} {\bibfnamefont {R.}~\bibnamefont {{Di Candia}}}, \bibinfo {author}
  {\bibfnamefont {Q.}~\bibnamefont {Chen}}, \bibinfo {author} {\bibfnamefont
  {Y.}~\bibnamefont {Nojiri}}, \bibinfo {author} {\bibfnamefont
  {K.}~\bibnamefont {Inomata}}, \bibinfo {author} {\bibfnamefont
  {Y.}~\bibnamefont {Nakamura}}, \bibinfo {author} {\bibfnamefont
  {M.}~\bibnamefont {Partanen}}, \bibinfo {author} {\bibfnamefont
  {A.}~\bibnamefont {Marx}}, \bibinfo {author} {\bibfnamefont {R.}~\bibnamefont
  {Gross}},\ and\ \bibinfo {author} {\bibfnamefont {F.}~\bibnamefont {Deppe}},\
  }\bibfield  {title} {\bibinfo {title} {\textit{Experimental quantum
  teleportation of propagating microwaves}},\ }\href
  {https://doi.org/10.1126/sciadv.abk0891} {\bibfield  {journal} {\bibinfo
  {journal} {Sci. Adv.}\ }\textbf {\bibinfo {volume} {\textbf{7}}},\ \bibinfo
  {pages} {eabk0891} (\bibinfo {year} {2021})}\BibitemShut {NoStop}%
\bibitem [{\citenamefont {Renger}\ \emph {et~al.}(2021)\citenamefont {Renger},
  \citenamefont {Pogorzalek}, \citenamefont {Chen}, \citenamefont {Nojiri},
  \citenamefont {Inomata}, \citenamefont {Nakamura}, \citenamefont {Partanen},
  \citenamefont {Marx}, \citenamefont {Gross}, \citenamefont {Deppe},\ and\
  \citenamefont {Fedorov}}]{Renger2021}%
  \BibitemOpen
  \bibfield  {author} {\bibinfo {author} {\bibfnamefont {M.}~\bibnamefont
  {Renger}}, \bibinfo {author} {\bibfnamefont {S.}~\bibnamefont {Pogorzalek}},
  \bibinfo {author} {\bibfnamefont {Q.}~\bibnamefont {Chen}}, \bibinfo {author}
  {\bibfnamefont {Y.}~\bibnamefont {Nojiri}}, \bibinfo {author} {\bibfnamefont
  {K.}~\bibnamefont {Inomata}}, \bibinfo {author} {\bibfnamefont
  {Y.}~\bibnamefont {Nakamura}}, \bibinfo {author} {\bibfnamefont
  {M.}~\bibnamefont {Partanen}}, \bibinfo {author} {\bibfnamefont
  {A.}~\bibnamefont {Marx}}, \bibinfo {author} {\bibfnamefont {R.}~\bibnamefont
  {Gross}}, \bibinfo {author} {\bibfnamefont {F.}~\bibnamefont {Deppe}},\ and\
  \bibinfo {author} {\bibfnamefont {K.~G.}\ \bibnamefont {Fedorov}},\
  }\bibfield  {title} {\bibinfo {title} {\textit{Beyond the standard quantum
  limit for parametric amplification of broadband signals}},\ }\href
  {https://doi.org/10.1038/s41534-021-00495-y} {\bibfield  {journal} {\bibinfo
  {journal} {npj Quantum Inf.}\ }\textbf {\bibinfo {volume} {\textbf{7}}},\
  \bibinfo {pages} {160} (\bibinfo {year} {2021})}\BibitemShut {NoStop}%
\bibitem [{\citenamefont {Lescanne}\ \emph {et~al.}(2019)\citenamefont
  {Lescanne}, \citenamefont {Verney}, \citenamefont {Ficheux}, \citenamefont
  {Devoret}, \citenamefont {Huard}, \citenamefont {Mirrahimi},\ and\
  \citenamefont {Leghtas}}]{Lescanne2019}%
  \BibitemOpen
  \bibfield  {author} {\bibinfo {author} {\bibfnamefont {R.}~\bibnamefont
  {Lescanne}}, \bibinfo {author} {\bibfnamefont {L.}~\bibnamefont {Verney}},
  \bibinfo {author} {\bibfnamefont {Q.}~\bibnamefont {Ficheux}}, \bibinfo
  {author} {\bibfnamefont {M.~H.}\ \bibnamefont {Devoret}}, \bibinfo {author}
  {\bibfnamefont {B.}~\bibnamefont {Huard}}, \bibinfo {author} {\bibfnamefont
  {M.}~\bibnamefont {Mirrahimi}},\ and\ \bibinfo {author} {\bibfnamefont
  {Z.}~\bibnamefont {Leghtas}},\ }\bibfield  {title} {\bibinfo {title}
  {\textit{Escape of a Driven Quantum Josephson Circuit into Unconfined
  States}},\ }\href {https://doi.org/10.1103/PhysRevApplied.11.014030}
  {\bibfield  {journal} {\bibinfo  {journal} {Phys. Rev. Appl.}\ }\textbf
  {\bibinfo {volume} {\textbf{11}}},\ \bibinfo {pages} {014030} (\bibinfo
  {year} {2019})}\BibitemShut {NoStop}%
\bibitem [{\citenamefont {Verney}\ \emph {et~al.}(2019)\citenamefont {Verney},
  \citenamefont {Lescanne}, \citenamefont {Devoret},\ and\ \citenamefont
  {Leghtas}}]{Verney2019}%
  \BibitemOpen
  \bibfield  {author} {\bibinfo {author} {\bibfnamefont {L.}~\bibnamefont
  {Verney}}, \bibinfo {author} {\bibfnamefont {R.}~\bibnamefont {Lescanne}},
  \bibinfo {author} {\bibfnamefont {M.~H.}\ \bibnamefont {Devoret}},\ and\
  \bibinfo {author} {\bibfnamefont {Z.}~\bibnamefont {Leghtas}},\ }\bibfield
  {title} {\bibinfo {title} {\textit{Structural Instability of Driven Josephson
  Circuits Prevented by an Inductive Shunt}},\ }\href
  {https://doi.org/10.1103/PhysRevApplied.11.024003} {\bibfield  {journal}
  {\bibinfo  {journal} {Phys. Rev. Appl.}\ }\textbf {\bibinfo {volume}
  {\textbf{10}}},\ \bibinfo {pages} {024003} (\bibinfo {year}
  {2019})}\BibitemShut {NoStop}%
\bibitem [{\citenamefont {Shillito}\ \emph {et~al.}(2022)\citenamefont
  {Shillito}, \citenamefont {Petrescu}, \citenamefont {Cohen}, \citenamefont
  {Beall}, \citenamefont {Hauru}, \citenamefont {Ganahl}, \citenamefont
  {Lewis}, \citenamefont {Vidal},\ and\ \citenamefont {Blais}}]{Shillito2022}%
  \BibitemOpen
  \bibfield  {author} {\bibinfo {author} {\bibfnamefont {R.}~\bibnamefont
  {Shillito}}, \bibinfo {author} {\bibfnamefont {A.}~\bibnamefont {Petrescu}},
  \bibinfo {author} {\bibfnamefont {J.}~\bibnamefont {Cohen}}, \bibinfo
  {author} {\bibfnamefont {J.}~\bibnamefont {Beall}}, \bibinfo {author}
  {\bibfnamefont {M.}~\bibnamefont {Hauru}}, \bibinfo {author} {\bibfnamefont
  {M.}~\bibnamefont {Ganahl}}, \bibinfo {author} {\bibfnamefont {A.~G.}\
  \bibnamefont {Lewis}}, \bibinfo {author} {\bibfnamefont {G.}~\bibnamefont
  {Vidal}},\ and\ \bibinfo {author} {\bibfnamefont {A.}~\bibnamefont {Blais}},\
  }\bibfield  {title} {\bibinfo {title} {\textit{Dynamics of Transmon
  Ionization}},\ }\href {https://doi.org/10.1103/PhysRevApplied.18.034031}
  {\bibfield  {journal} {\bibinfo  {journal} {Phys. Rev. Appl.}\ }\textbf
  {\bibinfo {volume} {\textbf{18}}},\ \bibinfo {pages} {034031} (\bibinfo
  {year} {2022})}\BibitemShut {NoStop}%
\bibitem [{\citenamefont {Cohen}\ \emph {et~al.}(2023)\citenamefont {Cohen},
  \citenamefont {Petrescu}, \citenamefont {Shillito},\ and\ \citenamefont
  {Blais}}]{Cohen2023}%
  \BibitemOpen
  \bibfield  {author} {\bibinfo {author} {\bibfnamefont {J.}~\bibnamefont
  {Cohen}}, \bibinfo {author} {\bibfnamefont {A.}~\bibnamefont {Petrescu}},
  \bibinfo {author} {\bibfnamefont {R.}~\bibnamefont {Shillito}},\ and\
  \bibinfo {author} {\bibfnamefont {A.}~\bibnamefont {Blais}},\ }\bibfield
  {title} {\bibinfo {title} {{Reminiscence of classical chaos in driven
  transmons}},\ }\href {https://doi.org/10.1103/PRXQuantum.4.020312} {\bibfield
   {journal} {\bibinfo  {journal} {PRX Quantum}\ }\textbf {\bibinfo {volume}
  {\textbf{4}}},\ \bibinfo {pages} {020312} (\bibinfo {year}
  {2023})}\BibitemShut {NoStop}%
\bibitem [{\citenamefont {Boissonneault}\ \emph {et~al.}(2010)\citenamefont
  {Boissonneault}, \citenamefont {Gambetta},\ and\ \citenamefont
  {Blais}}]{Boissonneault2010}%
  \BibitemOpen
  \bibfield  {author} {\bibinfo {author} {\bibfnamefont {M.}~\bibnamefont
  {Boissonneault}}, \bibinfo {author} {\bibfnamefont {J.~M.}\ \bibnamefont
  {Gambetta}},\ and\ \bibinfo {author} {\bibfnamefont {A.}~\bibnamefont
  {Blais}},\ }\bibfield  {title} {\bibinfo {title} {\textit{Improved
  superconducting qubit readout by qubit-induced nonlinearities}},\ }\href
  {https://doi.org/10.1103/PhysRevLett.105.100504} {\bibfield  {journal}
  {\bibinfo  {journal} {Phys. Rev. Lett.}\ }\textbf {\bibinfo {volume}
  {\textbf{105}}},\ \bibinfo {pages} {100504} (\bibinfo {year}
  {2010})}\BibitemShut {NoStop}%
\bibitem [{\citenamefont {Bishop}\ \emph {et~al.}(2010)\citenamefont {Bishop},
  \citenamefont {Ginossar},\ and\ \citenamefont {Girvin}}]{Bishop2010}%
  \BibitemOpen
  \bibfield  {author} {\bibinfo {author} {\bibfnamefont {L.~S.}\ \bibnamefont
  {Bishop}}, \bibinfo {author} {\bibfnamefont {E.}~\bibnamefont {Ginossar}},\
  and\ \bibinfo {author} {\bibfnamefont {S.~M.}\ \bibnamefont {Girvin}},\
  }\bibfield  {title} {\bibinfo {title} {\textit{Response of the Strongly
  Driven Jaynes-Cummings Oscillator}},\ }\href
  {https://doi.org/10.1103/PhysRevLett.105.100505} {\bibfield  {journal}
  {\bibinfo  {journal} {Phys. Rev. Lett.}\ }\textbf {\bibinfo {volume}
  {\textbf{105}}},\ \bibinfo {pages} {100505} (\bibinfo {year}
  {2010})}\BibitemShut {NoStop}%
\bibitem [{\citenamefont {Pietik{\"{a}}inen}\ \emph {et~al.}(2019)\citenamefont
  {Pietik{\"{a}}inen}, \citenamefont {Tuorila}, \citenamefont {Golubev},\ and\
  \citenamefont {Paraoanu}}]{Pietikainen2019}%
  \BibitemOpen
  \bibfield  {author} {\bibinfo {author} {\bibfnamefont {I.}~\bibnamefont
  {Pietik{\"{a}}inen}}, \bibinfo {author} {\bibfnamefont {J.}~\bibnamefont
  {Tuorila}}, \bibinfo {author} {\bibfnamefont {D.~S.}\ \bibnamefont
  {Golubev}},\ and\ \bibinfo {author} {\bibfnamefont {G.~S.}\ \bibnamefont
  {Paraoanu}},\ }\bibfield  {title} {\bibinfo {title} {\textit{Photon blockade
  and the quantum-to-classical transition in the driven-dissipative Josephson
  pendulum coupled to a resonator}},\ }\href
  {https://doi.org/10.1103/PhysRevA.99.063828} {\bibfield  {journal} {\bibinfo
  {journal} {Phys. Rev. A}\ }\textbf {\bibinfo {volume} {\textbf{99}}},\
  \bibinfo {pages} {063828} (\bibinfo {year} {2019})}\BibitemShut {NoStop}%
\bibitem [{\citenamefont {Carmichael}(2015)}]{Carmichael2015}%
  \BibitemOpen
  \bibfield  {author} {\bibinfo {author} {\bibfnamefont {H.~J.}\ \bibnamefont
  {Carmichael}},\ }\bibfield  {title} {\bibinfo {title} {\textit{Breakdown of
  photon blockade: A dissipative quantum phase transition in zero
  dimensions}},\ }\href {https://doi.org/10.1103/PhysRevX.5.031028} {\bibfield
  {journal} {\bibinfo  {journal} {Phys. Rev. X}\ }\textbf {\bibinfo {volume}
  {\textbf{5}}},\ \bibinfo {pages} {031028} (\bibinfo {year}
  {2015})}\BibitemShut {NoStop}%
\bibitem [{\citenamefont {Mavrogordatos}(2016)}]{Mavrogordatos2016}%
  \BibitemOpen
  \bibfield  {author} {\bibinfo {author} {\bibfnamefont {T.~K.}\ \bibnamefont
  {Mavrogordatos}},\ }\bibfield  {title} {\bibinfo {title} {\textit{Quantum
  phase transitions in the driven dissipative Jaynes-Cummings oscillator: From
  the dispersive regime to resonance}},\ }\href
  {https://doi.org/10.1209/0295-5075/116/54001} {\bibfield  {journal} {\bibinfo
   {journal} {EPL}\ }\textbf {\bibinfo {volume} {\textbf{116}}},\ \bibinfo
  {pages} {54001} (\bibinfo {year} {2016})}\BibitemShut {NoStop}%
\bibitem [{\citenamefont {Fink}\ \emph {et~al.}(2018)\citenamefont {Fink},
  \citenamefont {Schade}, \citenamefont {H{\"{o}}fling}, \citenamefont
  {Schneider},\ and\ \citenamefont {Imamoglu}}]{Fink2018}%
  \BibitemOpen
  \bibfield  {author} {\bibinfo {author} {\bibfnamefont {T.}~\bibnamefont
  {Fink}}, \bibinfo {author} {\bibfnamefont {A.}~\bibnamefont {Schade}},
  \bibinfo {author} {\bibfnamefont {S.}~\bibnamefont {H{\"{o}}fling}}, \bibinfo
  {author} {\bibfnamefont {C.}~\bibnamefont {Schneider}},\ and\ \bibinfo
  {author} {\bibfnamefont {A.}~\bibnamefont {Imamoglu}},\ }\bibfield  {title}
  {\bibinfo {title} {\textit{Signatures of a dissipative phase transition in
  photon correlation measurements}},\ }\href
  {https://doi.org/10.1038/s41567-017-0020-9} {\bibfield  {journal} {\bibinfo
  {journal} {Nat. Phys.}\ }\textbf {\bibinfo {volume} {\textbf{14}}},\ \bibinfo
  {pages} {365} (\bibinfo {year} {2018})}\BibitemShut {NoStop}%
\bibitem [{\citenamefont {Chen}\ \emph {et~al.}(2023)\citenamefont {Chen},
  \citenamefont {Fischer}, \citenamefont {Nojiri}, \citenamefont {Renger},
  \citenamefont {Xie}, \citenamefont {Partanen}, \citenamefont {Pogorzalek},
  \citenamefont {Fedorov}, \citenamefont {Marx}, \citenamefont {Deppe},\ and\
  \citenamefont {Gross}}]{Chen2023}%
  \BibitemOpen
  \bibfield  {author} {\bibinfo {author} {\bibfnamefont {Q.-M.}\ \bibnamefont
  {Chen}}, \bibinfo {author} {\bibfnamefont {M.}~\bibnamefont {Fischer}},
  \bibinfo {author} {\bibfnamefont {Y.}~\bibnamefont {Nojiri}}, \bibinfo
  {author} {\bibfnamefont {M.}~\bibnamefont {Renger}}, \bibinfo {author}
  {\bibfnamefont {E.}~\bibnamefont {Xie}}, \bibinfo {author} {\bibfnamefont
  {M.}~\bibnamefont {Partanen}}, \bibinfo {author} {\bibfnamefont
  {S.}~\bibnamefont {Pogorzalek}}, \bibinfo {author} {\bibfnamefont {K.~G.}\
  \bibnamefont {Fedorov}}, \bibinfo {author} {\bibfnamefont {A.}~\bibnamefont
  {Marx}}, \bibinfo {author} {\bibfnamefont {F.}~\bibnamefont {Deppe}},\ and\
  \bibinfo {author} {\bibfnamefont {R.}~\bibnamefont {Gross}},\ }\bibfield
  {title} {\bibinfo {title} {\textit{Quantum behavior of a superconducting
  Duffing oscillator at the dissipative phase transition}},\ }\href
  {https://doi.org/10.1038/s41467-023-38217-x} {\bibfield  {journal} {\bibinfo
  {journal} {Nat. Commun.}\ }\textbf {\bibinfo {volume} {\textbf{14}}},\
  \bibinfo {pages} {2896} (\bibinfo {year} {2023})}\BibitemShut {NoStop}%
\bibitem [{\citenamefont {Utermann}\ \emph {et~al.}(1994)\citenamefont
  {Utermann}, \citenamefont {Dittrich},\ and\ \citenamefont
  {H{\"{a}}nggi}}]{Utermann1994}%
  \BibitemOpen
  \bibfield  {author} {\bibinfo {author} {\bibfnamefont {R.}~\bibnamefont
  {Utermann}}, \bibinfo {author} {\bibfnamefont {T.}~\bibnamefont {Dittrich}},\
  and\ \bibinfo {author} {\bibfnamefont {P.}~\bibnamefont {H{\"{a}}nggi}},\
  }\bibfield  {title} {\bibinfo {title} {\textit{Tunneling and the onset of
  chaos in a driven bistable system}},\ }\href
  {https://doi.org/10.1103/PhysRevE.49.273} {\bibfield  {journal} {\bibinfo
  {journal} {Phys. Rev. E}\ }\textbf {\bibinfo {volume} {\textbf{49}}},\
  \bibinfo {pages} {273} (\bibinfo {year} {1994})}\BibitemShut {NoStop}%
\bibitem [{\citenamefont {Reed}\ \emph {et~al.}(2010)\citenamefont {Reed},
  \citenamefont {Dicarlo}, \citenamefont {Johnson}, \citenamefont {Sun},
  \citenamefont {Schuster}, \citenamefont {Frunzio},\ and\ \citenamefont
  {Schoelkopf}}]{Reed2010}%
  \BibitemOpen
  \bibfield  {author} {\bibinfo {author} {\bibfnamefont {M.~D.}\ \bibnamefont
  {Reed}}, \bibinfo {author} {\bibfnamefont {L.}~\bibnamefont {Dicarlo}},
  \bibinfo {author} {\bibfnamefont {B.~R.}\ \bibnamefont {Johnson}}, \bibinfo
  {author} {\bibfnamefont {L.}~\bibnamefont {Sun}}, \bibinfo {author}
  {\bibfnamefont {D.~I.}\ \bibnamefont {Schuster}}, \bibinfo {author}
  {\bibfnamefont {L.}~\bibnamefont {Frunzio}},\ and\ \bibinfo {author}
  {\bibfnamefont {R.~J.}\ \bibnamefont {Schoelkopf}},\ }\bibfield  {title}
  {\bibinfo {title} {\textit{High-fidelity readout in circuit quantum
  electrodynamics using the jaynes-cummings nonlinearity}},\ }\href
  {https://doi.org/10.1103/PhysRevLett.105.173601} {\bibfield  {journal}
  {\bibinfo  {journal} {Phys. Rev. Lett.}\ }\textbf {\bibinfo {volume}
  {\textbf{105}}},\ \bibinfo {pages} {173601} (\bibinfo {year}
  {2010})}\BibitemShut {NoStop}%
\bibitem [{\citenamefont {Gambetta}\ \emph {et~al.}(2008)\citenamefont
  {Gambetta}, \citenamefont {Blais}, \citenamefont {Boissonneault},
  \citenamefont {Houck}, \citenamefont {Schuster},\ and\ \citenamefont
  {Girvin}}]{Gambetta2008}%
  \BibitemOpen
  \bibfield  {author} {\bibinfo {author} {\bibfnamefont {J.}~\bibnamefont
  {Gambetta}}, \bibinfo {author} {\bibfnamefont {A.}~\bibnamefont {Blais}},
  \bibinfo {author} {\bibfnamefont {M.}~\bibnamefont {Boissonneault}}, \bibinfo
  {author} {\bibfnamefont {A.~A.}\ \bibnamefont {Houck}}, \bibinfo {author}
  {\bibfnamefont {D.~I.}\ \bibnamefont {Schuster}},\ and\ \bibinfo {author}
  {\bibfnamefont {S.~M.}\ \bibnamefont {Girvin}},\ }\bibfield  {title}
  {\bibinfo {title} {\textit{Quantum trajectory approach to circuit QED:
  Quantum jumps and the Zeno effect}},\ }\href
  {https://doi.org/10.1103/PhysRevA.77.012112} {\bibfield  {journal} {\bibinfo
  {journal} {Phys. Rev. A}\ }\textbf {\bibinfo {volume} {\textbf{77}}},\
  \bibinfo {pages} {012112} (\bibinfo {year} {2008})}\BibitemShut {NoStop}%
\bibitem [{\citenamefont {Krantz}\ \emph {et~al.}(2019)\citenamefont {Krantz},
  \citenamefont {Kjaergaard}, \citenamefont {Yan}, \citenamefont {Orlando},
  \citenamefont {Gustavsson},\ and\ \citenamefont {Oliver}}]{Krantz2019}%
  \BibitemOpen
  \bibfield  {author} {\bibinfo {author} {\bibfnamefont {P.}~\bibnamefont
  {Krantz}}, \bibinfo {author} {\bibfnamefont {M.}~\bibnamefont {Kjaergaard}},
  \bibinfo {author} {\bibfnamefont {F.}~\bibnamefont {Yan}}, \bibinfo {author}
  {\bibfnamefont {T.~P.}\ \bibnamefont {Orlando}}, \bibinfo {author}
  {\bibfnamefont {S.}~\bibnamefont {Gustavsson}},\ and\ \bibinfo {author}
  {\bibfnamefont {W.~D.}\ \bibnamefont {Oliver}},\ }\bibfield  {title}
  {\bibinfo {title} {\textit{A quantum engineer's guide to superconducting
  qubits}},\ }\href {https://doi.org/10.1063/1.5089550} {\bibfield  {journal}
  {\bibinfo  {journal} {Applied Physics Reviews}\ }\textbf {\bibinfo {volume}
  {\textbf{6}}},\ \bibinfo {pages} {1} (\bibinfo {year} {2019})}\BibitemShut
  {NoStop}%
\bibitem [{\citenamefont {Elliott}\ and\ \citenamefont
  {Ginossar}(2016)}]{Elliott2016}%
  \BibitemOpen
  \bibfield  {author} {\bibinfo {author} {\bibfnamefont {M.}~\bibnamefont
  {Elliott}}\ and\ \bibinfo {author} {\bibfnamefont {E.}~\bibnamefont
  {Ginossar}},\ }\bibfield  {title} {\bibinfo {title} {\textit{Applications of
  the Fokker-Planck equation in circuit quantum electrodynamics}},\ }\href
  {https://doi.org/10.1103/PhysRevA.94.043840} {\bibfield  {journal} {\bibinfo
  {journal} {Phys. Rev. A}\ }\textbf {\bibinfo {volume} {\textbf{94}}},\
  \bibinfo {pages} {043840} (\bibinfo {year} {2016})}\BibitemShut {NoStop}%
\bibitem [{\citenamefont {R{\'{e}}nyi}(1960)}]{Renyi1960}%
  \BibitemOpen
  \bibfield  {author} {\bibinfo {author} {\bibfnamefont {A.}~\bibnamefont
  {R{\'{e}}nyi}},\ }\bibfield  {title} {\bibinfo {title} {\textit{On measures
  of information and entropy}},\ }\href@noop {} {\bibfield  {journal} {\bibinfo
   {journal} {Proceedings of the fourth Berkeley Symposium on Mathematics,
  Statistics and Probability}\ }\textbf {\bibinfo {volume} {1}},\ \bibinfo
  {pages} {547} (\bibinfo {year} {1960})}\BibitemShut {NoStop}%
\bibitem [{\citenamefont {Jin}\ and\ \citenamefont {Korepin}(2004)}]{Jin2004}%
  \BibitemOpen
  \bibfield  {author} {\bibinfo {author} {\bibfnamefont {B.~Q.}\ \bibnamefont
  {Jin}}\ and\ \bibinfo {author} {\bibfnamefont {V.~E.}\ \bibnamefont
  {Korepin}},\ }\bibfield  {title} {\bibinfo {title} {\textit{Quantum spin
  chain, Toeplitz determinants and the Fisher-Hartwig conjecture}},\ }\href
  {https://doi.org/10.1023/B:JOSS.0000037230.37166.42} {\bibfield  {journal}
  {\bibinfo  {journal} {J. Stat. Phys.}\ }\textbf {\bibinfo {volume}
  {\textbf{116}}},\ \bibinfo {pages} {79} (\bibinfo {year} {2004})}\BibitemShut
  {NoStop}%
\bibitem [{\citenamefont {Qolibikloo}\ and\ \citenamefont
  {Ghodsi}(2019)}]{Qolibikloo2019}%
  \BibitemOpen
  \bibfield  {author} {\bibinfo {author} {\bibfnamefont {S.}~\bibnamefont
  {Qolibikloo}}\ and\ \bibinfo {author} {\bibfnamefont {A.}~\bibnamefont
  {Ghodsi}},\ }\bibfield  {title} {\bibinfo {title} {\textit{More on phase
  transition and R{\'{e}}nyi entropy}},\ }\href
  {https://doi.org/10.1140/epjc/s10052-019-6927-9} {\bibfield  {journal}
  {\bibinfo  {journal} {Eur. Phys. J. C}\ }\textbf {\bibinfo {volume}
  {\textbf{79}}},\ \bibinfo {pages} {406} (\bibinfo {year} {2019})}\BibitemShut
  {NoStop}%
\bibitem [{\citenamefont {Hines}\ \emph {et~al.}(2005)\citenamefont {Hines},
  \citenamefont {McKenzie},\ and\ \citenamefont {Milburn}}]{Hines2005}%
  \BibitemOpen
  \bibfield  {author} {\bibinfo {author} {\bibfnamefont {A.~P.}\ \bibnamefont
  {Hines}}, \bibinfo {author} {\bibfnamefont {R.~H.}\ \bibnamefont
  {McKenzie}},\ and\ \bibinfo {author} {\bibfnamefont {G.~J.}\ \bibnamefont
  {Milburn}},\ }\bibfield  {title} {\bibinfo {title} {\textit{Quantum
  entanglement and fixed-point bifurcations}},\ }\href
  {https://doi.org/10.1103/PhysRevA.71.042303} {\bibfield  {journal} {\bibinfo
  {journal} {Phys. Rev. A}\ }\textbf {\bibinfo {volume} {\textbf{71}}},\
  \bibinfo {pages} {042303} (\bibinfo {year} {2005})}\BibitemShut {NoStop}%
\bibitem [{\citenamefont {C{\'{o}}rcoles}\ \emph {et~al.}(2011)\citenamefont
  {C{\'{o}}rcoles}, \citenamefont {Chow}, \citenamefont {Gambetta},
  \citenamefont {Rigetti}, \citenamefont {Rozen}, \citenamefont {Keefe},
  \citenamefont {{Beth Rothwell}}, \citenamefont {Ketchen},\ and\ \citenamefont
  {Steffen}}]{Corcoles2011}%
  \BibitemOpen
  \bibfield  {author} {\bibinfo {author} {\bibfnamefont {A.~D.}\ \bibnamefont
  {C{\'{o}}rcoles}}, \bibinfo {author} {\bibfnamefont {J.~M.}\ \bibnamefont
  {Chow}}, \bibinfo {author} {\bibfnamefont {J.~M.}\ \bibnamefont {Gambetta}},
  \bibinfo {author} {\bibfnamefont {C.}~\bibnamefont {Rigetti}}, \bibinfo
  {author} {\bibfnamefont {J.~R.}\ \bibnamefont {Rozen}}, \bibinfo {author}
  {\bibfnamefont {G.~A.}\ \bibnamefont {Keefe}}, \bibinfo {author}
  {\bibfnamefont {M.}~\bibnamefont {{Beth Rothwell}}}, \bibinfo {author}
  {\bibfnamefont {M.~B.}\ \bibnamefont {Ketchen}},\ and\ \bibinfo {author}
  {\bibfnamefont {M.}~\bibnamefont {Steffen}},\ }\bibfield  {title} {\bibinfo
  {title} {\textit{Protecting superconducting qubits from radiation}},\ }\href
  {https://doi.org/10.1063/1.3658630} {\bibfield  {journal} {\bibinfo
  {journal} {Appl. Phys. Lett.}\ }\textbf {\bibinfo {volume} {\textbf{99}}},\
  \bibinfo {pages} {181906} (\bibinfo {year} {2011})}\BibitemShut {NoStop}%
\bibitem [{\citenamefont {Geerlings}\ \emph {et~al.}(2013)\citenamefont
  {Geerlings}, \citenamefont {Leghtas}, \citenamefont {Pop}, \citenamefont
  {Shankar}, \citenamefont {Frunzio}, \citenamefont {Schoelkopf}, \citenamefont
  {Mirrahimi},\ and\ \citenamefont {Devoret}}]{Geerlings2013}%
  \BibitemOpen
  \bibfield  {author} {\bibinfo {author} {\bibfnamefont {K.}~\bibnamefont
  {Geerlings}}, \bibinfo {author} {\bibfnamefont {Z.}~\bibnamefont {Leghtas}},
  \bibinfo {author} {\bibfnamefont {I.~M.}\ \bibnamefont {Pop}}, \bibinfo
  {author} {\bibfnamefont {S.}~\bibnamefont {Shankar}}, \bibinfo {author}
  {\bibfnamefont {L.}~\bibnamefont {Frunzio}}, \bibinfo {author} {\bibfnamefont
  {R.~J.}\ \bibnamefont {Schoelkopf}}, \bibinfo {author} {\bibfnamefont
  {M.}~\bibnamefont {Mirrahimi}},\ and\ \bibinfo {author} {\bibfnamefont
  {M.~H.}\ \bibnamefont {Devoret}},\ }\bibfield  {title} {\bibinfo {title}
  {\textit{Demonstrating a driven reset protocol for a superconducting
  qubit}},\ }\href {https://doi.org/10.1103/PhysRevLett.110.120501} {\bibfield
  {journal} {\bibinfo  {journal} {Phys. Rev. Lett.}\ }\textbf {\bibinfo
  {volume} {\textbf{110}}},\ \bibinfo {pages} {120501} (\bibinfo {year}
  {2013})}\BibitemShut {NoStop}%
\bibitem [{\citenamefont {Jin}\ \emph {et~al.}(2015)\citenamefont {Jin},
  \citenamefont {Kamal}, \citenamefont {Sears}, \citenamefont {Gudmundsen},
  \citenamefont {Hover}, \citenamefont {Miloshi}, \citenamefont {Slattery},
  \citenamefont {Yan}, \citenamefont {Yoder}, \citenamefont {Orlando},
  \citenamefont {Gustavsson},\ and\ \citenamefont {Oliver}}]{Jin2015}%
  \BibitemOpen
  \bibfield  {author} {\bibinfo {author} {\bibfnamefont {X.~Y.}\ \bibnamefont
  {Jin}}, \bibinfo {author} {\bibfnamefont {A.}~\bibnamefont {Kamal}}, \bibinfo
  {author} {\bibfnamefont {A.~P.}\ \bibnamefont {Sears}}, \bibinfo {author}
  {\bibfnamefont {T.}~\bibnamefont {Gudmundsen}}, \bibinfo {author}
  {\bibfnamefont {D.}~\bibnamefont {Hover}}, \bibinfo {author} {\bibfnamefont
  {J.}~\bibnamefont {Miloshi}}, \bibinfo {author} {\bibfnamefont
  {R.}~\bibnamefont {Slattery}}, \bibinfo {author} {\bibfnamefont
  {F.}~\bibnamefont {Yan}}, \bibinfo {author} {\bibfnamefont {J.}~\bibnamefont
  {Yoder}}, \bibinfo {author} {\bibfnamefont {T.~P.}\ \bibnamefont {Orlando}},
  \bibinfo {author} {\bibfnamefont {S.}~\bibnamefont {Gustavsson}},\ and\
  \bibinfo {author} {\bibfnamefont {W.~D.}\ \bibnamefont {Oliver}},\ }\bibfield
   {title} {\bibinfo {title} {\textit{Thermal and Residual Excited-State
  Population in a 3D Transmon Qubit}},\ }\href
  {https://doi.org/10.1103/PhysRevLett.114.240501} {\bibfield  {journal}
  {\bibinfo  {journal} {Phys. Rev. Lett.}\ }\textbf {\bibinfo {volume}
  {\textbf{114}}},\ \bibinfo {pages} {240501} (\bibinfo {year}
  {2015})}\BibitemShut {NoStop}%
\bibitem [{\citenamefont {Balembois}\ \emph {et~al.}()\citenamefont
  {Balembois}, \citenamefont {Travesedo}, \citenamefont {Pallegoix},
  \citenamefont {May}, \citenamefont {Billaud}, \citenamefont {Villiers},
  \citenamefont {Est{\`{e}}ve}, \citenamefont {Vion}, \citenamefont {Bertet},\
  and\ \citenamefont {Flurin}}]{Balembois2023}%
  \BibitemOpen
  \bibfield  {author} {\bibinfo {author} {\bibfnamefont {L.}~\bibnamefont
  {Balembois}}, \bibinfo {author} {\bibfnamefont {J.}~\bibnamefont
  {Travesedo}}, \bibinfo {author} {\bibfnamefont {L.}~\bibnamefont
  {Pallegoix}}, \bibinfo {author} {\bibfnamefont {A.}~\bibnamefont {May}},
  \bibinfo {author} {\bibfnamefont {E.}~\bibnamefont {Billaud}}, \bibinfo
  {author} {\bibfnamefont {M.}~\bibnamefont {Villiers}}, \bibinfo {author}
  {\bibfnamefont {D.}~\bibnamefont {Est{\`{e}}ve}}, \bibinfo {author}
  {\bibfnamefont {D.}~\bibnamefont {Vion}}, \bibinfo {author} {\bibfnamefont
  {P.}~\bibnamefont {Bertet}},\ and\ \bibinfo {author} {\bibfnamefont
  {E.}~\bibnamefont {Flurin}},\ }\href {http://arxiv.org/abs/2307.03614}
  {\bibinfo {title} {\textit{Practical Single Microwave Photon Counter with
  $10^\mathrm{-22}$ $\mathrm{W/\sqrt{Hz}}$ sensitivity}}},\ \Eprint
  {https://arxiv.org/abs/2307.03614} {arXiv:2307.03614} \BibitemShut {NoStop}%
\bibitem [{\citenamefont {Gambetta}\ \emph {et~al.}(2011)\citenamefont
  {Gambetta}, \citenamefont {Motzoi}, \citenamefont {Merkel},\ and\
  \citenamefont {Wilhelm}}]{Gambetta2011}%
  \BibitemOpen
  \bibfield  {author} {\bibinfo {author} {\bibfnamefont {J.~M.}\ \bibnamefont
  {Gambetta}}, \bibinfo {author} {\bibfnamefont {F.}~\bibnamefont {Motzoi}},
  \bibinfo {author} {\bibfnamefont {S.~T.}\ \bibnamefont {Merkel}},\ and\
  \bibinfo {author} {\bibfnamefont {F.~K.}\ \bibnamefont {Wilhelm}},\
  }\bibfield  {title} {\bibinfo {title} {\textit{Analytic control methods for
  high-fidelity unitary operations in a weakly nonlinear oscillator}},\ }\href
  {https://doi.org/10.1103/PhysRevA.83.012308} {\bibfield  {journal} {\bibinfo
  {journal} {Phys. Rev. A}\ }\textbf {\bibinfo {volume} {\textbf{83}}},\
  \bibinfo {pages} {012308} (\bibinfo {year} {2011})}\BibitemShut {NoStop}%
\bibitem [{\citenamefont {Chen}\ \emph {et~al.}(2016)\citenamefont {Chen},
  \citenamefont {Kelly}, \citenamefont {Quintana}, \citenamefont {Barends},
  \citenamefont {Campbell}, \citenamefont {Chen}, \citenamefont {Chiaro},
  \citenamefont {Dunsworth}, \citenamefont {Fowler}, \citenamefont {Lucero},
  \citenamefont {Jeffrey}, \citenamefont {Megrant}, \citenamefont {Mutus},
  \citenamefont {Neeley}, \citenamefont {Neill}, \citenamefont {O'Malley},
  \citenamefont {Roushan}, \citenamefont {Sank}, \citenamefont {Vainsencher},
  \citenamefont {Wenner}, \citenamefont {White}, \citenamefont {Korotkov},\
  and\ \citenamefont {Martinis}}]{Chen2016}%
  \BibitemOpen
  \bibfield  {author} {\bibinfo {author} {\bibfnamefont {Z.}~\bibnamefont
  {Chen}}, \bibinfo {author} {\bibfnamefont {J.}~\bibnamefont {Kelly}},
  \bibinfo {author} {\bibfnamefont {C.}~\bibnamefont {Quintana}}, \bibinfo
  {author} {\bibfnamefont {R.}~\bibnamefont {Barends}}, \bibinfo {author}
  {\bibfnamefont {B.}~\bibnamefont {Campbell}}, \bibinfo {author}
  {\bibfnamefont {Y.}~\bibnamefont {Chen}}, \bibinfo {author} {\bibfnamefont
  {B.}~\bibnamefont {Chiaro}}, \bibinfo {author} {\bibfnamefont
  {A.}~\bibnamefont {Dunsworth}}, \bibinfo {author} {\bibfnamefont {A.~G.}\
  \bibnamefont {Fowler}}, \bibinfo {author} {\bibfnamefont {E.}~\bibnamefont
  {Lucero}}, \bibinfo {author} {\bibfnamefont {E.}~\bibnamefont {Jeffrey}},
  \bibinfo {author} {\bibfnamefont {A.}~\bibnamefont {Megrant}}, \bibinfo
  {author} {\bibfnamefont {J.}~\bibnamefont {Mutus}}, \bibinfo {author}
  {\bibfnamefont {M.}~\bibnamefont {Neeley}}, \bibinfo {author} {\bibfnamefont
  {C.}~\bibnamefont {Neill}}, \bibinfo {author} {\bibfnamefont {P.~J.}\
  \bibnamefont {O'Malley}}, \bibinfo {author} {\bibfnamefont {P.}~\bibnamefont
  {Roushan}}, \bibinfo {author} {\bibfnamefont {D.}~\bibnamefont {Sank}},
  \bibinfo {author} {\bibfnamefont {A.}~\bibnamefont {Vainsencher}}, \bibinfo
  {author} {\bibfnamefont {J.}~\bibnamefont {Wenner}}, \bibinfo {author}
  {\bibfnamefont {T.~C.}\ \bibnamefont {White}}, \bibinfo {author}
  {\bibfnamefont {A.~N.}\ \bibnamefont {Korotkov}},\ and\ \bibinfo {author}
  {\bibfnamefont {J.~M.}\ \bibnamefont {Martinis}},\ }\bibfield  {title}
  {\bibinfo {title} {\textit{Measuring and Suppressing Quantum State Leakage in
  a Superconducting Qubit}},\ }\href
  {https://doi.org/10.1103/PhysRevLett.116.020501} {\bibfield  {journal}
  {\bibinfo  {journal} {Phys. Rev. Lett.}\ }\textbf {\bibinfo {volume}
  {\textbf{116}}},\ \bibinfo {pages} {020501} (\bibinfo {year}
  {2016})}\BibitemShut {NoStop}%
\bibitem [{\citenamefont {Yan}\ \emph {et~al.}(2016)\citenamefont {Yan},
  \citenamefont {Gustavsson}, \citenamefont {Kamal}, \citenamefont {Birenbaum},
  \citenamefont {Sears}, \citenamefont {Hover}, \citenamefont {Gudmundsen},
  \citenamefont {Rosenberg}, \citenamefont {Samach}, \citenamefont {Weber},
  \citenamefont {Yoder}, \citenamefont {Orlando}, \citenamefont {Clarke},
  \citenamefont {Kerman},\ and\ \citenamefont {Oliver}}]{Yan2016}%
  \BibitemOpen
  \bibfield  {author} {\bibinfo {author} {\bibfnamefont {F.}~\bibnamefont
  {Yan}}, \bibinfo {author} {\bibfnamefont {S.}~\bibnamefont {Gustavsson}},
  \bibinfo {author} {\bibfnamefont {A.}~\bibnamefont {Kamal}}, \bibinfo
  {author} {\bibfnamefont {J.}~\bibnamefont {Birenbaum}}, \bibinfo {author}
  {\bibfnamefont {A.~P.}\ \bibnamefont {Sears}}, \bibinfo {author}
  {\bibfnamefont {D.}~\bibnamefont {Hover}}, \bibinfo {author} {\bibfnamefont
  {T.~J.}\ \bibnamefont {Gudmundsen}}, \bibinfo {author} {\bibfnamefont
  {D.}~\bibnamefont {Rosenberg}}, \bibinfo {author} {\bibfnamefont
  {G.}~\bibnamefont {Samach}}, \bibinfo {author} {\bibfnamefont
  {S.}~\bibnamefont {Weber}}, \bibinfo {author} {\bibfnamefont {J.~L.}\
  \bibnamefont {Yoder}}, \bibinfo {author} {\bibfnamefont {T.~P.}\ \bibnamefont
  {Orlando}}, \bibinfo {author} {\bibfnamefont {J.}~\bibnamefont {Clarke}},
  \bibinfo {author} {\bibfnamefont {A.~J.}\ \bibnamefont {Kerman}},\ and\
  \bibinfo {author} {\bibfnamefont {W.~D.}\ \bibnamefont {Oliver}},\ }\bibfield
   {title} {\bibinfo {title} {\textit{The flux qubit revisited to enhance
  coherence and reproducibility}},\ }\href
  {https://doi.org/10.1038/ncomms12964} {\bibfield  {journal} {\bibinfo
  {journal} {Nat. Commun.}\ }\textbf {\bibinfo {volume} {\textbf{7}}},\
  \bibinfo {pages} {12964} (\bibinfo {year} {2016})}\BibitemShut {NoStop}%
\end{thebibliography}%
\end{document}